\documentclass[aps,twocolumn,superscriptaddress,amsmath,amssymb,showkeys,floatfix]{revtex4-1}

\usepackage[a4paper, left=.55in, right=.55in, top=.7in, bottom=.7in]{geometry}

\usepackage{times}

\usepackage{bm}
\usepackage[utf8]{inputenc}
\usepackage{braket}
\usepackage{amssymb,amsmath}
\usepackage[british]{babel}
\usepackage[autostyle=true]{csquotes}
\usepackage{enumitem}
\usepackage[colorlinks=true,citecolor=blue,linkcolor=blue,urlcolor=blue]{hyperref}
\usepackage{cleveref}
\usepackage{amsthm}
\usepackage{mathtools}

\usepackage[export]{adjustbox}

\newtheorem*{EPT}{Equitable Partition Theorem}
\newtheorem*{nEPT}{Nonequitable Partition Theorem}
\newtheorem*{Remark}{Remark}

\crefname{Remark}{remark\!}{Remark\!}
\crefname{equation}{Eq.\!}{Eqs.\!}
\crefname{figure}{Fig.\!}{Figs.\!}
\crefname{chapter}{Chap.\!}{Chaps.\!}
\crefname{section}{Sec.\!}{Secs.\!}
\crefname{appendix}{App.\!}{Apps.\!}

\usepackage{graphicx}
\graphicspath{ {./} }

\newcommand{\symmetryOp}{\mathcal{S}}
\newcommand{\FF}{\mathbb{F}}
\newcommand{\TT}{\mathbb{T}}
\renewcommand{\SS}{\mathbb{S}}

\let\originalleft\left
\let\originalright\right
\renewcommand{\left}{\mathopen{}\mathclose\bgroup\originalleft}
\renewcommand{\right}{\aftergroup\egroup\originalright}

\makeatletter
\renewcommand*\env@matrix[1][\arraystretch]{%
	\edef\arraystretch{#1}%
	\hskip -\arraycolsep
	\let\@ifnextchar\new@ifnextchar
	\array{*\c@MaxMatrixCols c}}
\makeatother

\makeatletter
\appto{\appendix}{%
  \@ifstar{\def\theequation@prefix{A.}}%
          {}%
}
\makeatother

\begin{document}


\title{Compact localized states and flat bands from local symmetry partitioning}

\author{M. R\"ontgen}
\affiliation{%
Zentrum f\"ur optische Quantentechnologien, Universit\"at Hamburg, Luruper Chaussee 149, 22761 Hamburg, Germany
}%

\author{C. V. Morfonios}
\affiliation{%
Zentrum f\"ur optische Quantentechnologien, Universit\"at Hamburg, Luruper Chaussee 149, 22761 Hamburg, Germany
}%

\author{P. Schmelcher}
\affiliation{%
Zentrum f\"ur optische Quantentechnologien, Universit\"at Hamburg, Luruper Chaussee 149, 22761 Hamburg, Germany
}%
\affiliation{%
The Hamburg Centre for Ultrafast Imaging, Universit\"at Hamburg, Luruper Chaussee 149, 22761 Hamburg, Germany
}%


\begin{abstract}
\begin{center}
 Journal reference: \href{https://journals.aps.org/prb/abstract/10.1103/PhysRevB.97.035161}{Phys. Rev. B \textbf{97}, 035161 (2018)}
\end{center}

We propose a framework for the connection between local symmetries of discrete Hamiltonians and the design of compact localized states.
Such compact localized states are used for the creation of tunable, local symmetry-induced bound states in an energy continuum and flat energy bands for periodically repeated local symmetries in one- and two-dimensional lattices.
The framework is based on very recent theorems in graph theory which are here employed to obtain a block partitioning of the Hamiltonian induced by the symmetry of a given system under local site permutations.
The diagonalization of the Hamiltonian is thereby reduced to finding the eigenspectra of smaller matrices, with eigenvectors automatically divided into compact localized and extended states.
We distinguish between local symmetry operations which commute with the Hamiltonian, and those which do not commute due to an asymmetric coupling to the surrounding sites.
While valuable as a computational tool for versatile discrete systems with locally symmetric structures, the approach provides in particular a unified, intuitive, and efficient route to the flexible design of compact localized states at desired energies.
\end{abstract}


\maketitle

\section{\label{sec:Introduction}Introduction}

Compact localized states \cite{Flach2014_EPL_105_30001_DetanglingFlat,Maimaiti2017_PRB_95_115135_CompactLocalized}, i.\,e. wave excitations that strictly vanish outside a finite subpart of a system, are caused by destructive interference in the presence of local spatial symmetries \cite{Flach2014_EPL_105_30001_DetanglingFlat}. 
Contrary to the case of Anderson localization \cite{Anderson1958_PR_109_1492_AbsenceDiffusion}, where exponentially localized states are caused by disorder, compact localized states (CLSs) typically occur in perfectly ordered systems \cite{Flach2014_EPL_105_30001_DetanglingFlat}.
They were early deduced from symmetry principles in bipartite lattices \cite{Sutherland1986_PRB_34_5208-5211_Localizationelectronicwave}, and studied more recently in, e.\,g., frustrated hopping models \cite{Bergman2008_PRB_78_125104_BandTouching} as well as magnonic \cite{Derzhko2006_EPJB-_52_23_UniversalLow-temperature} and interacting \cite{Derzhko2010_PRB_81_014421_Low-temperatureProperties} systems.
A possible application of CLSs lies in information transmission \cite{
Vicencio2014_JO_16_015706_Diffraction-freeImage,
Rojas-Rojas2017_PRA_96_043803_Quantumlocalizedstates,
Xia2016_OL_41_1435_DemonstrationFlat-band} 
and directly stems from their compactness:
Being an eigenstate of the Hamiltonian, a CLS does not spread out spatially during evolution, while it is much less challenging to excite than a regular extended eigenstate.
For example, CLSs are ideal candidates for the transmission of information along photonic waveguide arrays avoiding `crosstalk' between waveguides \cite{Vicencio2015_PRL_114_245503_ObservationLocalized}. 
Further, CLSs essentially enable the appearance of isolated bound states within a scattering continuum \cite{
Vonneuman1929_PZ_30_467_UberMerkwuerdige,
Stillinger1975_PRA_11_446_BoundStates,
Hsu2016_NRM_1_16048_Boundstatescontinuum}. 
Such states were, e.\,g., realized recently as a symmetry-induced topological eigenstate subspace of coupled-chain setups \cite{Xiao2017_PRL_118_166803_TopologicalSubspaceInducedBound}.
On a computational level, CLSs induced by symmetries may also be used as a symmetry-adapted basis for numerical computations \cite{Klein2015_CMCC_74_247_LocalSymmetries}. 
In periodic lattice systems, macroscopically degenerate CLSs lead to the occurrence of flat, i.\,e. dispersionless, energy bands  \cite{Bodyfelt2014_PRL_113_236403_FlatbandsUnder}.
Flat bands are studied in different contexts, including the quantum Hall effect in topologically non-trivial lattices \cite{Parameswaran2013_CRP_14_816_FractionalQuantum,Yang2012_PRB_86_241112_TopologicalFlat,Neupert2011_PRL_106_236804_FractionalQuantum,Tang2011_PRL_106_236802_High-temperatureFractional}, induced metal-insulator transitions \cite{Souza2009_PRB_79_153104_Flat-bandLocalization,Danieli2015_PRB_91_235134_Flat-bandEngineering} and non-Hermitian quantum mechanics \cite{Leykam2017_PRB_96_064305_FlatBands,Ramezani2017_PRA_96_011802_Non-hermiticity-induFlat}.

Different approaches have been suggested to \emph{design} systems featuring CLSs and flat bands. 
They are based on strategies such as so-called `origami rules' \cite{Dias2015_SR_5_srep16852_OrigamiRules}, the repetition of mini-arrays \cite{Morales-inostroza2016_PRA_94_043831_SimpleMethod}, working on bipartite Hamiltonians \cite{Ramachandran2017_PRB_96_161104_Chiralflatbands}, detangling the lattice into Fano lattices \cite{Flach2014_EPL_105_30001_DetanglingFlat} or even more general approaches, such as band engineering \cite{Xu2015_SR_5_srep18181_DesignFull-k-space} or generator principles \cite{Maimaiti2017_PRB_95_115135_CompactLocalized}. 
Most of these works are based on the presence of different kinds of \emph{local symmetries}, i.e. on the invariance of a subset of matrix elements under a site permutation. 
In general, local symmetries of the underlying Hamiltonian are indirectly encoded into its eigenstates, as has been demonstrated recently in various contexts \cite{Kalozoumis2013_PRA_87_032113_LocalSymmetries,Kalozoumis2013_PRA_88_033857_LocalSymmetries,Kalozoumis2014_PRL_113_050403_InvariantsBroken,Kalozoumis2015_AP_362_684_InvariantCurrents,Zampetakis2016_JPAMT_49_195304_InvariantCurrent,Morfonios2017_AP_385_623_NonlocalDiscrete,Roentgen2017_AP_380_135_Non-localCurrents,Schmelcher2017_JCP_146_044116_DynamicsLocal,Wulf2016_PRE_93_052215_ExposingLocal}.
However, not every locally symmetric system features CLSs, and a systematic framework linking a theory of local symmetries to the formation and control of both CLSs and the resulting flat bands is still missing.

In the present work, we take a step in this direction by applying very recent graph theoretical results to generic single-particle discrete Hamiltonians. 
The resulting unifying framework connects two types of local symmetries to the occurrence of CLSs, flat bands, and bound states in the continuum. 
Complementing many of the above CLS design strategies, this framework uniquely pairs a high degree of control with an in-depth understanding of the impact of local symmetries.
Technically, we apply and generalize two recently published theorems \cite{Klein2015_CMCC_74_247_LocalSymmetries,Barrett2017_LAA_513_409_EquitableDecompositions,Francis2017_LAA_532_432_ExtensionsApplications,Fritscher2016_SJMAA_37_260_ExploringSymmetries} to general Hamiltonian matrices.
These theorems, which we refer to as the equitable and nonequitable partition theorems, quantify the effect of certain local symmetries of the Hamiltonian matrix $H$ underlying a given discrete system.
Specifically, the equitable partition theorem (EPT) applies to locally acting symmetry transformations which commute with $H$, while the nonequitable partition theorem (nEPT) applies to a subclass of transformations that do not commute with $H$.
In essence, the theorems assert a symmetry-induced decomposition of $H$ into a direct sum (i.\,e. block-diagonal form) of smaller matrices, whose spectrum and eigenvectors thereby determine those of $H$.
In particular, the eigenvectors of submatrices corresponding to symmetric subsystems of the complete setup uniquely provide all existing CLSs of $H$ together with their eigenenergies.
The remaining submatrix is analogously connected to extended eigenstates (non-CLSs) of $H$.

In the context of periodic lattices, the presence of local symmetries is thus shown to \emph{automatically} enforce the presence of flat bands, while the (n)EPT can be used to control both the flat \emph{and} dispersive bands of the system.
The approach can be seen as complementary to the general and powerful design principle of Refs.\,\cite{Flach2014_EPL_105_30001_DetanglingFlat,Maimaiti2017_PRB_95_115135_CompactLocalized}  based on elementwise conditions on the underlying eigenvalue equation, in that it solely relies on generalized symmetry concepts.
Moreover, the methodology can be used to reduce the computational effort of diagonalisation by exploiting local symmetries present in the Hamiltonian.

We apply the framework to the design of both flat bands and symmetry-induced bound states in the continuum.
It should be emphasized that the approach allows for the design of symmetry-induced flat bands at prescribed energies in \emph{arbitrary} dimensions. 
Moreover, since it is solely based on the symmetries of a complex-valued square matrix, the framework is applicable to a broad range of physical problems, treated by e.g. multichannel scattering theory or dyadic Green functions \cite{Rusek2000_PRA_61_022704_RandomGreen,Pinheiro2004_PRE_69_026605_ProbingAnderson,Christofi2016_OL_41_1933_ProbingScattering}.
We thus believe that this work may inspire the exploration of the effect of local symmetries in the broader research community.

The paper is structured as follows. 
\Cref{sec:subsymmetries} introduces the concept and description of local symmetries and subsequently states the EPT and nEPT in terms of simple example setups.
In \cref{sec:applications} we demonstrate the methodology in the design of bound states in the continuum and flat band lattices.
\Cref{sec:conclusions} contains our conclusions.

\section{Local symmetries and equitable partitions} \label{sec:subsymmetries}

The setting we will operate on is the eigenvalue problem 
\begin{equation} \label{eq:evp}
 H \bm{\phi} = E \bm{\phi}
\end{equation}
of a Hamiltonian matrix $H$ modeling a (lattice) system of sites $n$ with elements 
\begin{equation} \label{eq:discreteHamiltonian}
H_{mn} = \begin{cases} v_{n}, & m = n \\ h_{m,n} \ne 0, & n \in \mathcal{N}(m)  \\ 0, & \mbox{else}\end{cases}
\end{equation}
where $\mathcal{N}(n)$ denotes a set of neighboring sites connected to site $n$ via a non-vanishing hopping.
$H$ is graphically represented by a (weighted) graph with vertices connected by edges for corresponding nonzero hoppings, like in \cref{fig:comsubsym}. 
Throughout, we will use different vertex sizes and coloring to indicate different values of the onsite potential of the represented Hamiltonian.
The considered model can be seen as a generalized tight-binding network, with more than just next-neighbor hopping being allowed. 
Such a model is extensively used to describe single-electron phenomena, such as e.\,g. localization in lattice systems \cite{Chalker2010_PRB_82_104209_AndersonLocalization,Bergman2008_PRB_78_125104_BandTouching}.
It also effectively describes, for instance, arrays of evanescently coupled photonic waveguides, in terms of which both flat bands \cite{Mukherjee2015_PRL_114_245504_ObservationLocalized,Vicencio2015_PRL_114_245503_ObservationLocalized,Doretto2015_PRB_92_245124_Flat-bandFerromagnetism} and bound states in the continuum \cite{Plotnik2011_PRL_107_183901_ExperimentalObservation,Zhang2013_PRA_87_023613_BoundStates} have been studied.

\subsection{Commutative local symmetries} \label{sec:EPT}

\begin{figure}[t]
	\centering
	\includegraphics[max size={.7\columnwidth}{0.3\textheight}]{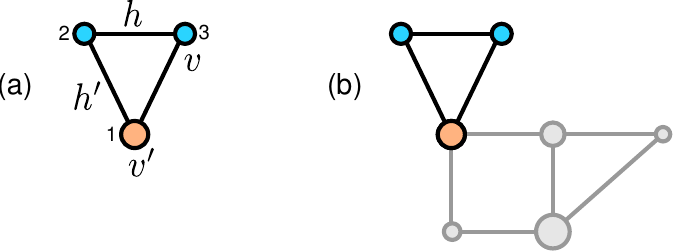}
	\caption{
	{\bf (a)} The Hamiltonian of a three-site system is represented by a graph with connected vertices, with vertex sizes (and colors) indicating different onsite potential values. 
	The system is symmetric under the permutation of sites $2$ and $3$, or globally symmetric under a left-right flip. 
	In {\bf (b)} the system is extended by attaching an arbitrary subsystem (grey) to site $1$ (which is fixed under the permutation), so that the original global symmetry becomes a commutative local symmetry.
	Independently of the parameters of the attached subsystem, the eigenvalue $v - h$ corresponding to a compact localized eigenstate on the two sites $2,3$ is always present in the Hamiltonian spectrum.
	}
	\label{fig:comsubsym}
\end{figure}

To introduce the concept of local symmetry, let us first consider the three-site system depicted in \cref{fig:comsubsym}\,(a).
Its Hamiltonian $H$ is invariant under permutation of sites $2$ and $3$, which represents a global left-right flip of the system. 
Since the corresponding permutation matrix $\varPi$ 
squares to unity ($\varPi^2 = I$) and commutes with $H$, their common eigenvectors will have definite parity under this permutation.
The spectrum 
$\sigma(H) = \{E_1,E_2,E_3\}$ of $H$ is given by 
$E_1 = v-h$ and 
$E_{2,3} = \frac{1}{2} [ v + v' + h \pm \sqrt{8 h'^2 + (h +v -v')^2}\,]$. The 
corresponding (unnormalized) eigenvectors are 
$\bm{\phi}^1 = [0,1,-1]^\top$ and 
$\bm{\phi}^{2,3} = [a_\pm,1,1]^\top$ (with $a_\pm$ depending on all system parameters), 
which indeed are of odd and even parity under $\varPi$, respectively.

Let us now connect an arbitrary subsystem to site $1$, still leaving the resulting composite system symmetric under the site permutation $2 \leftrightarrow 3$, as shown in the example of \Cref{fig:comsubsym}\,(b).
The corresponding permutation matrix $\varPi$ now has the dimension of the enlarged system, but performs the left-right flip only \emph{locally} on subsystem $\{1,2,3\}$, leaving site $1$ and the added subsystem identical, or \emph{fixed} under $\varPi$.
Since this local permutation commutes with the Hamiltonian, $\varPi H = H \varPi$, we say that the system possesses a \emph{commutative local symmetry}.

With $\varPi^2 = I$, the composite system eigenvectors again possess a definite parity under $\varPi$.
In particular, any eigenvector with odd parity will have zero amplitude on all sites fixed under $\varPi$ (since ${\phi}_n = -{\phi}_n$ for those sites), that is, the state is compactly localized on the symmetric subsystem.
Consequently, also the eigenvalues of these odd parity states are left unaltered by variations of the parameters (onsite and hopping elements) in the fixed subsystem.
In the example of \cref{fig:comsubsym}, the same eigenvalue $E_1 = v-h$ (of the odd parity eigenstate) will always be present in $\sigma(H)$, irrespectively of the fixed subsystem connected to site $1$. 
The corresponding eigenstate is localized only on sites $2$ and $3$ with opposite sign.

The above symmetry considerations, explaining the persistence of compact localized eigenstates of odd parity in the presence of commutative local symmetries, is formalized within graph theory by the so-called \emph{equitable partition theorem} (EPT) \cite{Trudeau1994____IntroductionGraph}, which also provides the eigenvalues of associated even parity eigenvectors. 
The term `equitable' denotes a partitioning of the vertices of a graph into non-overlapping classes such that for \emph{distinct} classes $A_{i}, A_{j}$ all vertices belonging to $A_{i}$ have the same number of adjacent vertices belonging to class $A_{j}$. 
Although this concept is limited to unweighted graphs, which can be represented by very specific matrices such as the adjacency or Laplacian matrix, the above definition of equitable partitions has recently been extended to general complex square matrices \cite{Barrett2017_LAA_513_409_EquitableDecompositions,Thuene2016_AC___ExploitingEquitable} including the model Hamiltonians considered here. 
In this generalization, a matrix is `equitably partitionable' if it can be partitioned into blocks of constant row sum. 
For instance, in \cref{fig:comsubsym}\,(a) this is the case as
\begin{equation}
	H =
	\left[
	\begin{array}{c|c c}
	v' & h' & h' \\
	\hline
	h' & v & h\\
	h' & h & v
	\end{array}
	\right]
\end{equation}
with row sums $v',2 h', h', v+h$.

\begin{figure}[t]
	\centering
	\includegraphics[max size={1\linewidth}{0.8\textheight}]{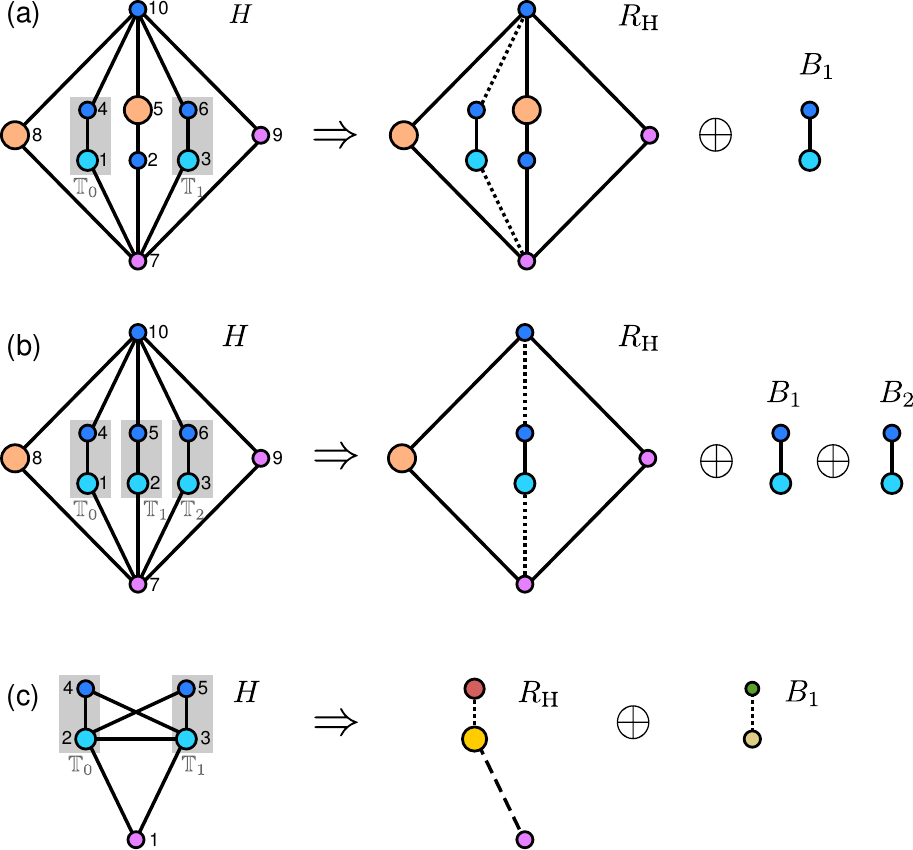}
	\caption{
	{\bf Left}: 
	Graphically represented Hamiltonian $H$ (with uniform hoppings $h$ and onsite elements indicated by different vertex sizes and coloring) of a system with local symmetry under mutual exchange of 
	{\bf (a)} two subparts $\TT_0$, $\TT_1$, 
	{\bf (b)} three subparts $\TT_0$, $\TT_1$, $\TT_2$, and 
	{\bf (c)} two interconnected subparts $\TT_0$, $\TT_1$ (indicated by gray background).
	{\bf Right}: 
	Using the equitable partition theorem (EPT), the Hamiltonian matrix $H$  is transformed ($\Rightarrow$) into a direct sum ($\oplus$) of the graphically represented matrices $R_{\text H}$ and $B_j$; \cref{eq:HBlockDiagonal,eq:hermitianDivisorMatrix}.
	In (a) and (b) there is no connection between the $\TT_i$, and only the divisor matrix $R_{\text H}$ has altered hoppings (dotted lines; $\sqrt{2}h$ in (a) and $\sqrt{3}h$ in (b)) compared to $H$.
	In (c), the intraconnections between $\TT_{0}$ and $\TT_{1}$ lead to altered onsite and hopping elements in both $R_{\text H}$ and $B_j$.
	}
	\label{fig:ept}
\end{figure}

Before stating the EPT, let us introduce the employed nomenclature through a suitable example. 
Consider the Hamiltonian graphically represented on the left hand side of \cref{fig:ept}\,(a).
The structure possesses a commutative local symmetry which can be visualized as a local flip of the sites $\{ 1,3\}$ and $\{4,6 \}$ around the axis running through the sites $7-2-5-10$. 
By `local' we mean that only the sites $\{1,3,4,6\}$ are flipped, while all other sites are unaffected.
This commutative local symmetry can be expressed by the commutation of the Hamiltonian with the permutation matrix
\begin{equation} \label{eq:permutationMatrixForm}
\varPi_{\symmetryOp} = 
\begin{bmatrix}
J_3 & 0 & 0 \\
0 & J_3 & 0 \\
0 & 0 & I_3
\end{bmatrix}
\equiv
J_3 \oplus J_3 \oplus I_3,
\end{equation}
$I_N$ and $J_N$ being the $N$-dimensional identity and exchange (antidiagonal, reverse identity) matrix, where $\oplus$ denotes direct sum (i.\,e. block-diagonal concatenation).
This symmetry operation on $H$, or \emph{automorphism} of its graph, can be described as a simultaneous permutation
\begin{align}
	\symmetryOp : 1 \mapsto 3,\; 3 \mapsto 1,\; 4 \mapsto 6,\; 6 \mapsto 4,
\end{align}
with all other sites being unaffected. This permutation $\symmetryOp: \{1,\ldots,10 \}\rightarrow \{1,\ldots 10 \}$ is commonly written in the so-called cyclic notation
\begin{equation} \label{eq:permuationOfFirstExample}
\symmetryOp = (2)(5)(7)(8)(9)(10)(1,3)(4,6) .
\end{equation}
Each tuple within parantheses in \cref{eq:permuationOfFirstExample} is called an \emph{orbit}. 
Orbits are classified by their size, i.\,e. by the number of sites they are comprised of. 
Orbits of size $1$ are called \emph{trivial}. 
Note that since permutations are bijective, orbits are always non-overlapping.

In accordance with the above, we will call a permutation $\symmetryOp: \{1,\ldots,N \}\rightarrow \{1,\ldots N \}$ satisfying
\begin{equation} \label{eq:defComPermutSym}
	H_{i,j} = H_{\symmetryOp(i),\symmetryOp(j)} \; \forall \; i,j \; \Leftrightarrow [H,\varPi_{\symmetryOp}] = 0
\end{equation}
a commutative local symmetry of $H$, with $\symmetryOp$ acting non-trivially on a subset of the system's sites, and $\varPi_{\symmetryOp}$ being the matrix representation of $\symmetryOp$.
If $\symmetryOp$ is a commutative local symmetry and all of its non-trivial orbits are of uniform size $k$, then we call it a \emph{basic commutative local symmetry of order $k$}. 
In the present example, $\symmetryOp$ given in \cref{eq:permuationOfFirstExample} is a basic commutative local symmetry of $H$ of order $2$ with two non-trivial and six trivial orbits.
It is clear from \cref{eq:defComPermutSym} that, in order to be a commutative local symmetry, a given permutation must leave the connections between sites invariant. 
For example, for $H$ in \cref{fig:ept}\,(a), the permutation 
\begin{equation} \label{eq:exampleNonCommutingSymmetry}
	\symmetryOp = (1)(2)(3)(4)(6)(7)(9)(10)(5,8)
\end{equation}
is not a commutative local symmetry:
While indeed $v_{8} = v_{5}$, $\symmetryOp$ breaks the connection, e.\,g., between sites $7$ (which is fixed under $\symmetryOp$) and $8$, $h_{\symmetryOp(8),\symmetryOp(7)} = 0 \ne h_{8,7}$, thus violating \cref{eq:defComPermutSym}.

If $H$ is represented graphically, commutative local symmetries of order $2$ can be seen as the invariance of the Hamiltonian under a local flip of a subsystem about an axis (which depends on how $H$ is depicted graphically), represented by a corresponding local permutation matrix $\varPi_\symmetryOp$. 
While this procedure aids in the graphical identification of commutative local symmetries, the notion of orbits is more powerful as it makes the description more compact in the case of increased local symmetry.
For example, in \cref{fig:ept}\,(b) all of the following are commutative local symmetries of $H$ of order $2$:
\begin{align} \label{eq:partialS}
\symmetryOp_a &= (7)(8)(9)(10)(2)(5)(4,6)(1,3) \nonumber \\
\symmetryOp_b &= (7)(8)(9)(10)(3)(6)(4,5)(1,2) \\
\symmetryOp_c &= (7)(8)(9)(10)(1)(4)(5,6)(2,3), \nonumber
\end{align}
each one corresponding to a local flip of a symmetric subsystem.
Those different local symmetries of order $2$ can now be unified into a single one of order $3$,
\begin{equation} \label{eq:basicAutomorphismOrder3}
\symmetryOp = (7)(8)(9)(10)(4,5,6)(1,2,3),
\end{equation}
i.\,e. by the simultaneous \emph{cyclic} permutations $1 \mapsto 2 \mapsto 3 \mapsto 1$ and $4 \mapsto 5 \mapsto 6 \mapsto 4$, exploiting the full local symmetry of the system at once. 
For the purpose of the EPT, $\symmetryOp$ is preferably chosen to be of highest possible order.

There is a fundamental connection between a basic commutative local symmetry $\symmetryOp$ of order ${{k}}$ and the structure of $H$: 
If it exists, then the sites of the system can be reordered \cite{Francis2017_LAA_532_432_ExtensionsApplications} by a suitable permutation $P$ such that $H$ is transformed into
\begin{equation} \label{eq:reorderingOfHamiltonian}
	\tilde{H} = P^{-1} H P = 
	\begin{bmatrix}
	F &  G & G & \dots & G\\
	G^{\dagger} & C_{0} & C_{1} & \dots & C_{{{k}}-1}\\
	G^{\dagger} & C_{{{k}}-1} & C_{0} & \ddots & \vdots \\
	\vdots & \vdots & \ddots & \ddots & C_{1} \\
	G^{\dagger} & C_{1} & \dots & C_{{{k}}-1} & C_{0}
	\end{bmatrix}
\end{equation}
with ${{k}}$ copies of the block $C_{0}$ on its diagonal and $C_{i} \in \mathbb{C}^{l \times l}$, where $l$ is the number of non-trivial orbits of $\symmetryOp$.
For a Hermitian Hamiltonian $H = H^{\dagger}$ the relation $C_{i} = C_{{{k}}-i}^{\dagger}$ holds.
A general procedure to transform $H$ to $\tilde{H}$ for a given basic commutative local symmetry $\symmetryOp$ of order $k$, with $f$ trivial and ${{l}}$ non-trivial orbits, is as follows: 
\begin{itemize}
	\item[(i)] Collect all $f$ sites fixed by $\symmetryOp$ into the subset $\FF$ of the set $\mathbb{N}$ of all sites.
	\item[(ii)] Construct a set $\TT_{0}$ of size $l$ by picking one arbitrary site from each one of the $l$ non-trivial orbits of $\symmetryOp$.
	\item[(iii)] Construct the sets $\TT_{i} = \symmetryOp^i \TT_{0}$, $i = 1,\ldots,{{k}}-1$, by the $i$-fold application of $\symmetryOp$ onto $\TT_{0}$ (noting that $\TT_{{{k}}} = \symmetryOp^k \TT_{0} = \TT_{0}$).
 	\item[(iv)] Construct $\tilde{H}$ in the form of \cref{eq:reorderingOfHamiltonian} using
	\begin{equation} \label{eq:DefinitionsOfSubmatrices}
	F = H_{\FF,\FF}, \;\; G = H_{\FF,\TT_{0}},\;\; C_{i} = H_{\TT_{0},\TT_{i}},
	\end{equation}
	where $H_{\mathbb{A},\mathbb{B}}$ denotes all elements $H_{mn}$ with $m \in \mathbb{A}$, $n \in \mathbb{B}$.
\end{itemize}
As an example, for the system in \cref{fig:ept}\,(a) we could choose $\TT_{0} = \{1,4\}$, so that $\TT_{1} = \{3,6\}$, and get
\begin{equation} \label{eq:firstReorderingC}
C_{0}  = 
\begin{bmatrix}
v_{1} & h_{1,4} \\
h_{4,1} & v_{4}
\end{bmatrix},\;\;
C_{1} = 
\begin{bmatrix}
0 & 0\\
0 & 0
\end{bmatrix},
\end{equation}
where $C_{1}$ vanishes since there are no inter-connections between $\TT_{0}$ and $\TT_{1}$. The other matrices are given as
\begin{equation} \label{eq:firstReorderingF}
F = \begin{bmatrix}
v_{2} & h_{2,5} & 0 & h_{2,8} & 0 & 0 \\
h_{5,2} & v_{5} & h_{5,7} & 0 & 0 & 0 \\
0 & h_{7,5} & v_{7} & 0 & h_{7,9} & h_{7,10} \\
h_{8,2} & 0 & 0 & v_{8} & h_{8,9} & h_{8,10} \\
0 & 0 & h_{9,7} & h_{9,8} & v_{9} & 0 \\
0 & 0 & h_{10,7} & h_{10,8} & 0 & v_{10}
\end{bmatrix},\; \;
G = \begin{bmatrix}
0 & 0\\
0 & 0\\
h_{7,1} & 0\\
0 & 0\\
0 & 0\\
0 & h_{10,4}
\end{bmatrix}.
\end{equation}
Note that $F$ constitutes the Hamiltonian of the isolated fixed subsystem $\FF$, the matrix (${{k}}=2$)
\begin{equation} \label{eq:symmetricConstituentsExample}
C = \begin{bmatrix}
C_{0} & C_{1} & \dots & C_{{{k}}-1} \\
C_{{{k}}-1} & C_{0} & \ddots & \vdots \\
\vdots & \ddots & \ddots & C_{1} \\
C_{1} & \dots & C_{{{k}}-1} & C_{0}
\end{bmatrix}
=
\begin{bmatrix}
v_{1} & h_{1,4} & 0 & 0 \\
h_{4,1} & v_{4} & 0 & 0 \\
0 & 0 & v_{1} & h_{1,4} \\
0 & 0 & h_{4,1} & v_{4}
\end{bmatrix}
\end{equation}
represents the isolated symmetric subsystem $\mathbb{S} = \mathbb{N} \smallsetminus \FF$ (which in this case are two uncoupled symmetric blocks), while $G$ couples the subsystems $\FF$ and $\mathbb{S}$. 
Thus, $\tilde{H}$ in \cref{eq:reorderingOfHamiltonian} can be seen as a symmetry-adapted restructuring of the Hamiltonian.

We can now, following Refs.\,\cite{Barrett2017_LAA_513_409_EquitableDecompositions,Francis2017_LAA_532_432_ExtensionsApplications}, state the

\begin{EPT} \label{theorem:equitablePartitionTheorem}
Let $H \in \mathbb{C}^{N\times N}$ have a commutative local symmetry $\symmetryOp$ of order $k$ with $l$ non-trivial and $f$ trivial orbits.
Then the following properties hold:

	\begin{enumerate}[label=\textbf{P\arabic*}]
	
	\item There exists an invertible, non-unitary matrix $M$ such that
	\begin{equation} \label{eq:HBlockDiagonal}
		H' = M^{-1} H M =
		R \oplus \bigoplus_{j = 1}^{k-1} B_j
		=
		\begin{bmatrix}
		R & 0 & \dots & 0\\
		0 & B_{1} & \ddots & \vdots \\
		\vdots & \ddots & \ddots & 0 \\
		0 & \dots & 0 & B_{k-1}
		\end{bmatrix}
	\end{equation}
	where
	\begin{equation} \label{eq:DivisorMatrix}
	R = 
	\begin{bmatrix}
	F & k \cdot G\\
	G^{\dagger} & B_{0}
	\end{bmatrix},~~
	B_{j} = \sum_{m=0}^{k-1} \omega^{j m} C_{m},
	\end{equation}
	with $\omega = e^{2 \pi i /k}$ and the matrices $F,G,C_m$ as defined in \cref{eq:DefinitionsOfSubmatrices}.
	
	\item \label{point:evsDecomposition} 
	The spectrum $\sigma(H)$ is given by
	\begin{equation} \label{eq:spectrumDecomposition}
	\sigma(H) =  \sigma(H') = \sigma(R) \cup \sigma(B_{1}) \cup \ldots \cup \sigma(B_{k-1}) .
	\end{equation}
	(regarding the hermiticity of $R$, see the remark below).

	\item \label{point:esDecomposition} The $N = f + kl$ eigenstates of the index-reordered matrix $\tilde{H}$ defined in \cref{eq:reorderingOfHamiltonian} are given by 
	\begin{equation} \label{eq:eigenstateDecomposition}
		\bm{\phi}^\nu =
		\begin{bmatrix}
		\bm{w}_\nu \\ \bm{v}_\nu \\ \bm{v}_\nu \\ \vdots \\ \bm{v}_\nu
		\end{bmatrix}, ~
		\bm{\phi}^{f+ml+r} =
		\begin{bmatrix}
		\bm{0}_{f} \\
		\bm{u}_{m,r} \\
		\omega^{m} \bm{u}_{m,r}\\
		\vdots \\
		\omega^{(k-1)m} \bm{u}_{m,r}\\
		\end{bmatrix},
	\end{equation}
	for $\nu \in [1,f+l]$ and $m \in [1,k-1], r \in [1,l]$, where
	$R {\footnotesize \begin{bmatrix}[.8] \bm{w}_\nu \\ \bm{v}_\nu \end{bmatrix}} = \lambda_\nu 
	{\footnotesize \begin{bmatrix}[.8] \bm{w}_\nu \\ \bm{v}_\nu \end{bmatrix}}$ with
	$\bm{w}_\nu \in \mathbb{C}^{f \times 1}$, $\bm{v}_\nu \in \mathbb{C}^{l \times 1}$, and 
	$B_{m} \bm{u}_{m,r} = \lambda_{m,r} \bm{u}_{m,r}$.
	The vectors $\bm{\phi}^{f+ml+r}$ are thereby compact localized on $\mathbb{S}$.

	\item \label{point:esSymmetry} 
	The first $f + l$ eigenvectors of $H$ are symmetric under $\symmetryOp$, while the remaining $(k-1)l$ eigenvectors are both compact localized and not symmetric under $\symmetryOp$. 
	Specifically, defining the index-reordered permutation matrix $\tilde{\varPi}_\symmetryOp = P^{-1} \tilde{\varPi}_\symmetryOp P$ with $P$ defined from \cref{eq:reorderingOfHamiltonian}, we have $\tilde{\varPi}_\symmetryOp \bm{\phi}^\nu = \bm{\phi}^\nu$ for $\nu \in [1,f+l]$, while the remaining compact $\mathbb{S}$-localized eigenvectors transform as
	\begin{equation}
	\tilde{\varPi}_\symmetryOp
	\begin{bmatrix}
	\bm{0}_f \\
	\bm{u}_{m,r} \\
	\omega^{m} \bm{u}_{m,r}\\
	\vdots \\
	\omega^{(k-1)m} \bm{u}_{m,r}\\
	\end{bmatrix} = 
	\begin{bmatrix}
	\bm{0}_f \\
	\omega^{(k-1)m} \bm{u}_{m,r} \\
	\bm{u}_{m,r}\\
	\vdots \\
	\omega^{(k-2)m} \bm{u}_{m,r}\\
	\end{bmatrix}.
	\end{equation}

	\end{enumerate}

\end{EPT}

\begin{Remark}
	The generally non-Hermitian `divisor' matrix $R$ defined in \cref{eq:DivisorMatrix} is isospectral to the similar Hermitian matrix
	\begin{equation} \label{eq:hermitianDivisorMatrix}
	R_{\text H} = 
	\begin{bmatrix}
	F & \sqrt{k} \cdot \, G \\
	\sqrt{k} \cdot \, G^{\dagger} & B_{0}
	\end{bmatrix} 
	= KRK^{-1}
	\end{equation}
	with eigenvectors ${\footnotesize \begin{bmatrix}[.8] \bm{w}_\nu \\ \bm{v}_\nu/\sqrt{k} \end{bmatrix}}$, where $K = I_f \oplus \sqrt{k} \cdot I_l$. 
	Thus, properties \ref{point:evsDecomposition} and \ref{point:esDecomposition} of the EPT hold if we replace $R$ by $R_{\text{H}}$ and $\bm{v}_\nu$ by $\bm{v}_\nu/\sqrt{k}$; 
	we shall do so in the remainder of this work.
\end{Remark}
We see that, in essence, the EPT uses the symmetries described by $\symmetryOp$ to acquire partial information from $H$, namely its spectral composition and corresponding eigenvector localization, without diagonalizing it. 
This information could indeed alternatively be obtained by considering the system's symmetry under local flip operations (represented by involutory matrices $\varPi$), as explained above.
In particular, however, the EPT provides \emph{all} eigenvalues and -vectors of $H$ in terms of those of the symmetry-adapted matrices $R$ and $B_j$, i.\,e. not only those of the `decoupled' CLSs.
Since $R$ and $B_j$ are of reduced dimension, the EPT may additionally offer a computational advantage in diagonalizing Hamiltonians of extended systems with commutative local symmetries.

\begin{figure*}[t!]
	\centering
	\includegraphics[max size={0.95 \linewidth}]{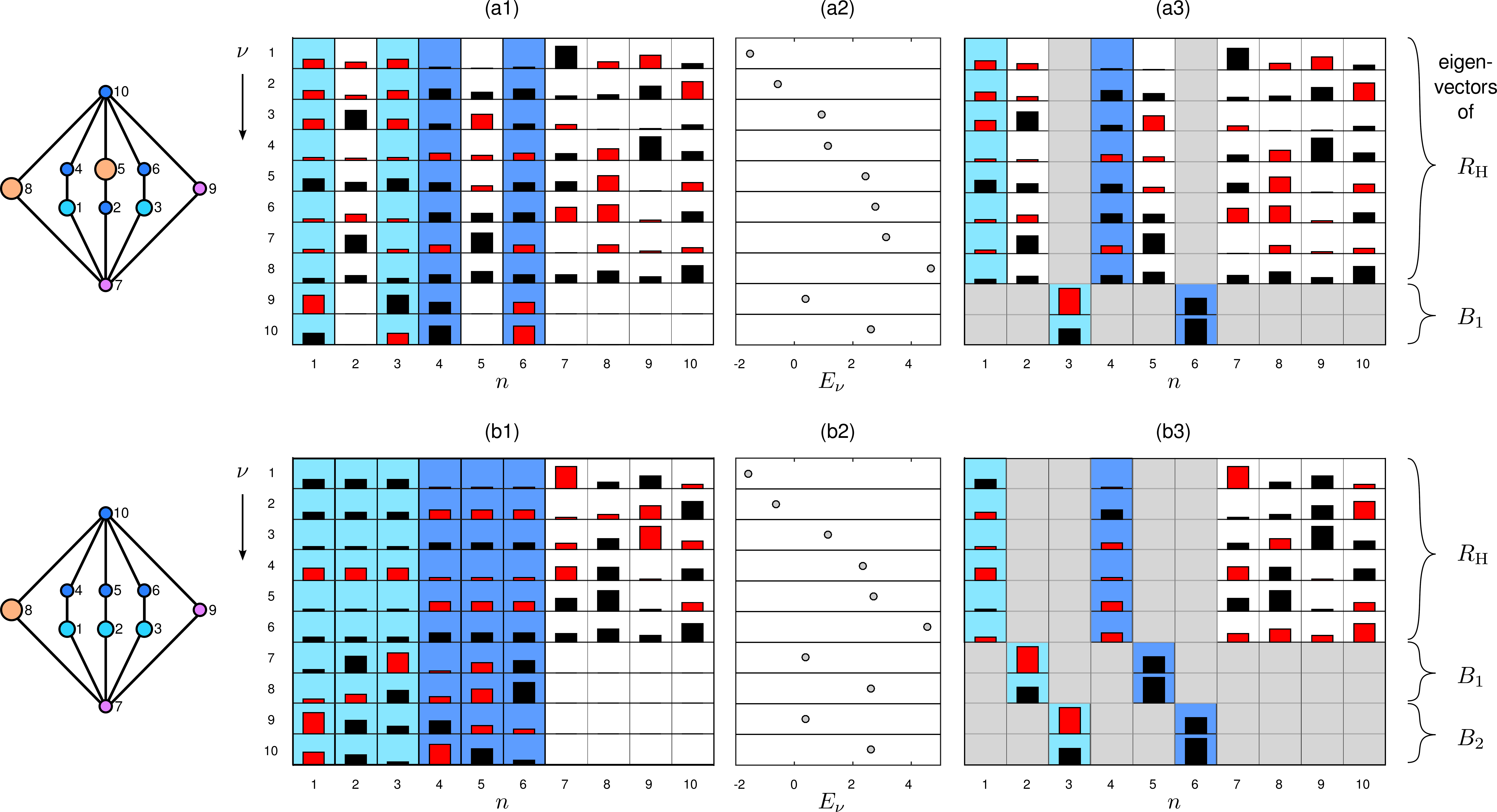}
	\caption{
	{\bf (a1)} Eigenvectors $\bm{\phi}^\nu$ of the Hamiltonian matrix $H$, with index $\nu$ ordered according to \cref{eq:eigenstateDecomposition} in the EPT, 
	{\bf (a2)} their corresponding eigenvalues $\{E_\nu\}$, and
	{\bf (a3)} eigenvectors of the matrices $R_{\text{H}}$ and $B_j$ ($j = 1,2$),
	for the system of \cref{fig:ept}\,(a), depicted on the left, with indicated onsite elements and homogeneous hoppings $h_{mn} = 1$.
	{\bf (b1)--(b3)} Similarly, but for the system of \cref{fig:ept}\,(b).
	The norm $|{\phi}^\nu_n|$ of each real eigenstate at each site $n$ is plotted in black (red) for ${\phi}^\nu_n > 0$ (${\phi}^\nu_n < 0$).
	The sites comprising the locally symmetric part of $H$ are indicated by corresponding light and dark blue background. 
	The eigenvectors of $R_{\text{H}}$ and $B_{j}$, which share eigenvalues $E_\nu$ with $H$, are spatially plotted following the site-indexing of $H$ with gray background for sites they are not defined on. 
	Note that the CLSs ($\nu = 9,10$ in (a1) and $\nu = 7,8,9,10$ in (b1)) are constructed from the components of $B_j$ eigenvectors at the same energy (with pairwise degeneracy for $\nu = 7,9$ and $8,10$ in (b1)), and that the remaining eigenstates are symmetric within the locally symmetric part. 
	}
	\label{fig:spectra}
\end{figure*}

To give a concrete impression of the EPT, we consider again the Hamiltonian $H$ in \cref{fig:ept}\,(a), which is transformed to the direct sum of matrices $R_{\text{H}}$ and $B_{1}$ according to \cref{eq:HBlockDiagonal}.
Recall that the similarity transformation involved preserves the spectrum of $H$, while the final block-diagonal form ensures property \ref{point:evsDecomposition} in the EPT.
The eigenvectors of $H$, $R_{\text{H}}$, and $B_{1}$ of \cref{fig:ept}\,(a) are shown in \cref{fig:spectra}\,(a1)-(a3) together with their eigenvalues. 
As predicted by the EPT (here with $f=6$, $k=2$, and $l=2$), there are two (antisymmetric) CLSs of $H$ (states $\nu = 9,10$) localized on the sites $\{ 1,3,4,6 \}$ that form a commutative local symmetry under the permutation $\symmetryOp$ of \cref{eq:permuationOfFirstExample}, while all other eigenstates are extended and symmetric under $\symmetryOp$.
In particular, the CLSs are constructed from the components of the eigenvectors of $B_{1}$ in \cref{fig:spectra}\,(a3).

The matrices $C_0,C_1$ and $F,G$, used in this example to construct the matrices $R$ and $B_j$ of the transformed Hamiltonian $H'$ in \cref{eq:HBlockDiagonal}, are given in \cref{eq:firstReorderingC,eq:firstReorderingF}, respectively, for the choice $\TT_{0} = \{1,4\}$ as initial orbit sites.
Note that the choice of $\TT_{0}$ generally affects the matrices $G,B_{1},\ldots, B_{k-1}$ (but not $B_{0}$), though does not change the resulting decomposition of the spectrum and the eigenvectors of the Hamiltonian. 
In the present example, the sites in each orbit are disconnected, so that $C_1$ vanishes and 
\begin{equation}
	B_{0} = B_{1} = C_{0} = 
	\begin{bmatrix}
	v_{1} & h_{1,4} \\
	h_{4,1} & v_{4}
	\end{bmatrix}
\end{equation}
from \cref{eq:DivisorMatrix} becomes the single submatrix corresponding to the two CLSs.

The EPT works in a completely similar form for the example in \cref{fig:ept}\,(b):
There are now $2$ orbits of size $k = 3$, leading to $(k-1)l = 4$ (pairwise degenerate) CLSs, as seen in \cref{fig:spectra}\,(b1)--(b2). 
Note that any two degenerate real CLS eigenvectors can be linearly combined to either be antisymmetric under one of the (partial) local symmetry transformations in \cref{eq:partialS}, or to be of the complex form in \cref{eq:eigenstateDecomposition}.
Also in this example there are no intra-orbit connections, and so we have $B_{0} = B_{1} = B_{2} = C_{0}$.

In contrast, the system shown in \cref{fig:ept}\,(c) is invariant under the permutation $\symmetryOp = (1)(2,3)(4,5)$ but has intraconnected orbits (or inter-connected local symmetry units $\TT_0$ and $\TT_1$), since $h_{2,3}, h_{2,5}, h_{4,3} \neq 0$.
In such a case the matrices $B_j$ differ;
here we have (with the choice $\TT_{0} = \{2,4\}$)
\begin{equation}
	R = \begin{bmatrix}
	v_{1}
	\end{bmatrix}, ~~
	B_{\substack{ \vspace{.1ex} \\ 0 \\ 1}} = \begin{bmatrix}
	v_{2} & h_{2,4} \\
	h_{4,2} & v_{4}
	\end{bmatrix} \pm \begin{bmatrix}
	h_{2,3} & h_{2,5} \\
	h_{4,3} & h_{4,5}
	\end{bmatrix}.
\end{equation}
Notably, the $B_{0}$ and $B_{1}$ here are given by adding and subtracting the intra-orbit connection, respectively. 
In \cref{sec:SymmetryAndDisorder} we will use this property to tailor periodic systems featuring bound states in the continuum.

\subsection{Noncommutative local symmetries} \label{sec:asymmetricEPT}

So far we have considered the case of local symmetries which, although localized within a part of a composite system, are represented by a permutation matrix $\varPi_{\symmetryOp}$ that commutes with the system Hamiltonian $H$.
We now show, partially following the procedure in Ref.\,\cite{Fritscher2016_SJMAA_37_260_ExploringSymmetries}, how the merits of the EPT can be extended to cases where a symmetric subsystem is \emph{asymmetrically} coupled to the rest of the system under the given site permutation.
Since $\varPi_{\symmetryOp}$ then does not commute with $H$, we call the underlying permutation $\symmetryOp$ a \emph{noncommutative local symmetry}. 
In the following, we will impose two further restrictions on $\symmetryOp$.

\begin{figure}[t]
	\centering
	\includegraphics[max size={0.75 \linewidth}{0.2 \textheight}]{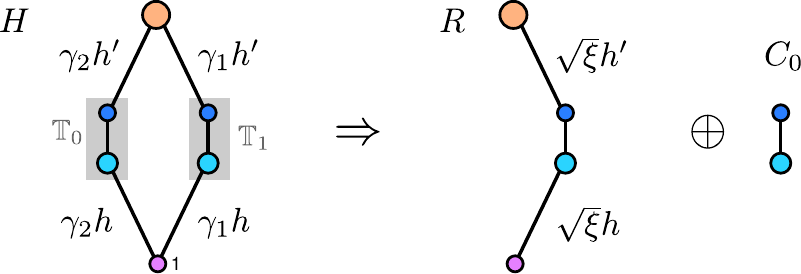}
	\caption{
	Graphically represented example Hamiltonian $H$ of the form in \cref{eq:SIAMHamiltonian} with a restricted noncommutative (local) symmetry under the exchange of subparts $\TT_0$ and $\TT_1$, and its partitioning into matrices $R$ and $C_0$ according to the nEPT with $\xi = \gamma_{1}^2 + \gamma_{2}^2$.
	}
	\label{fig:aEPT}
\end{figure}

Specifically, consider a Hamiltonian $H$ which can be index-reordered, in analogy to \cref{eq:reorderingOfHamiltonian}, into the form
\begin{equation} \label{eq:SIAMHamiltonian}
\tilde{H} =	\begin{bmatrix}
F &  \gamma_{1}^* G & \gamma_{2}^*G & \dots & \gamma_{k}^* G\\
\gamma_{1} G^{\dagger} & C_{0} & 0 & \dots & 0\\
\gamma_{2} G^{\dagger} & 0 & C_{0} & \ddots & \vdots \\
\vdots & \vdots & \ddots & \ddots & 0 \\
\gamma_{k} G^{\dagger} & 0 & \dots & 0 & C_{0}
\end{bmatrix}
\end{equation}
with generally complex parameters $\gamma_{1},\ldots,\gamma_{k}$.
Like in \cref{eq:reorderingOfHamiltonian}, the $k$ copies of $C_{0} \in \mathbb{C}^{l\times l}$ correspond to the same local symmetry units under permutation $\symmetryOp$, the matrix $F \in \mathbb{C}^{n\times n}$ corresponds to sites fixed by $\symmetryOp$, while $G \in \mathbb{C}^{n\times l}$ connects fixed to local symmetry sites. 
Now, however: 
(i) the local symmetry units are not interconnected (i.\,e. $C_{j > 0} = 0$), and 
(ii) while each of them is geometrically coupled to the fixed part $\FF$ in the same manner, the coupling strength for each unit is weighted by a factor $\gamma_{i}$.
Thus, if $\gamma_i \neq \gamma_j$ for some $i\ne j$, the coupling of $\SS$ (denoting the locally symmetric subsytems as a whole) to $\FF$ (denoting the fixed subsystem) is asymmetric, and $\symmetryOp$ is no longer a commutative local symmetry of $H$. 
A simple example is given by the system in \cref{fig:aEPT}; 
also by \cref{fig:ept}\,(a) if, e.\,g., only $h_{1,7}$ and $h_{4,10}$ were multiplied by a factor $\gamma$, or similarly in \cref{fig:ept}\,(b)---though not in \cref{fig:ept}\,(c), where there is local symmetry unit interconnection.

In the following, we will call noncommutative local symmetries fulfilling the above restrictions (i) and (ii) \emph{restricted noncommutative} ones. 
For such local symmetries the EPT can be modified, along the lines of Ref.\,\cite{Fritscher2016_SJMAA_37_260_ExploringSymmetries}, to the following

\begin{nEPT}
	Let $\tilde{H} \in \mathbb{C}^{N\times N}$ be of the form in \cref{eq:SIAMHamiltonian}, with $F \in \mathbb{C}^{f \times f}$, $k$ copies of $C_{0} \in \mathbb{C}^{l \times l}$ and $\tilde{\symmetryOp}$ a restricted noncommutative local symmetry of $\tilde{H}$.
	Then the following properties hold:

	\begin{enumerate}[label=\textbf{P\arabic*}]
		
		\item The eigenvalue spectrum of $\tilde{H}$ is given by $\sigma(\tilde{H}) = \sigma(H') = \sigma(R) \cup \sigma_{k-1}(C_0)$, where $H'$ is a similarity transform of $\tilde{H}$ and is given by
		\begin{equation} \label{eq:HBlockDiagonalSIAM}
		H' = 
		R \oplus \bigoplus_{m = 1}^{k-1} C_0, ~~
		R = 
		\begin{bmatrix}
		F & \sqrt{\xi} \cdot G\\
		\sqrt{\xi} \cdot G^{\dagger} & C_{0}
		\end{bmatrix}
		\end{equation}
		with $\xi = \sum_{j=1}^{k} \gamma_{j}^{2}$ and $\sigma_{i}(R)$ denoting $i$ copies of $\sigma(R)$ (i.\,e. $i$-fold degeneracy of those eigenvalues).
		
		\item \label{point:esDecompositionnEPT} The $N = f + kl$ eigenstates of $\tilde{H}$ are given by 
		\begin{equation} \label{eq:eigenstateDecompositionSIAMSecondClass}
		\bm{\phi}^\nu =
		\begin{bmatrix}[1.5]
		\bm{w}_\nu \\
		\frac{\gamma_{1}}{\sqrt{\xi}} \bm{v}_\nu \\
		\frac{\gamma_{2}}{\sqrt{\xi}}\bm{v}_\nu \\
		\vdots \\
		\frac{\gamma_{k}}{\sqrt{\xi}}\bm{v}_\nu
		\end{bmatrix},~
		\bm{\phi}^{f+ml+r} =
		\begin{bmatrix}[1.5]
		\bm{0}_{f} \\
		\frac{\gamma_{1}}{\gamma_{1}} \bm{u}_{0,r} \\
		\frac{\gamma_{2}}{\gamma_{1}} \bm{u}_{0,r}\\
		\vdots \\
		\frac{\gamma_{m}}{\gamma_{1}} \bm{u}_{0,r}\\
		-\frac{\sum_{i=1}^{m} \gamma_{i}^{2} }{\gamma_{1} \gamma_{m+1}} \bm{u}_{0,r}\\
		\bm{0}_{l} \\
		\vdots \\
		\bm{0}_{l} 
		\end{bmatrix}			
		\begin{matrix}
				  \vphantom{\begin{pmatrix}
					  \bm{0}_{f} \\
			  \frac{\gamma_{1}}{\gamma_{1}} \bm{u}_{0,r} \\
			  \frac{\gamma_{2}}{\gamma_{1}} \bm{u}_{0,r}\\
			  \vdots \\
			  \frac{\gamma_{m}}{\gamma_{1}} \bm{u}_{0,r}\\
			  -\frac{\sum_{i=1}^{m} \gamma_{i}^{2} }{\gamma_{1} \gamma_{m+1}}
					  \end{pmatrix} } \\
				  \vphantom{\vdots} \\
				  \left. \vphantom{\begin{pmatrix}
					  \bm{0}_{l} \\
					  \vdots \\
					  \bm{0}_{l} 
					  \end{pmatrix} } \right\} (k-1)-m \\
		\end{matrix},
		\end{equation}
		for $\nu \in [1,f+l]$, $m \in [1,k-1]$, and $r \in [1,l]$, where
		$R 
		{\footnotesize \begin{bmatrix}[.8] \bm{w}_\nu \\ \bm{v}_\nu \end{bmatrix}} = \lambda_\nu 
		{\footnotesize \begin{bmatrix}[.8] \bm{w}_\nu \\ \bm{v}_\nu \end{bmatrix}}$ with
		$\bm{w}_\nu \in \mathbb{C}^{f \times 1}$, $\bm{v}_\nu \in \mathbb{C}^{l \times 1}$, and 
		$C_{0} \bm{u}_{m,r} = \lambda_{0,r} \bm{u}_{m,r}$.

	\end{enumerate}
\end{nEPT}

The naming of the theorem was chosen to reflect the fact that, for unequal $\gamma_{i}$, the matrix (\ref{eq:SIAMHamiltonian}) cannot generally be partitioned into blocks with blockwise constant row sum;
that is, the matrix is `nonequitably' partitionable according to the definition above in \cref{sec:EPT}.
The theorem is proven in Ref.\,\cite{Fritscher2016_SJMAA_37_260_ExploringSymmetries} for real $\tilde{H}$, but is generalized here in a straightforward manner to complex $\tilde{H}$ and $\gamma_j$; see Appendix.
This may allow for the possibility to include appropriately applied external magnetic fields in the present symmetry-adapted construction of CLSs (via Peierls phase factors in the hopping elements \cite{Ramachandran2017_PRB_96_161104_Chiralflatbands}), or to include parametric gain and loss (via complex onsite elements \cite{Leykam2017_PRB_96_064305_FlatBands}).

It should here be mentioned that there exists a large class of local symmetries which are neither commutative nor restricted noncommutative.
Also, the restrictions for the nonequitable partition theorem (nEPT) to apply are indeed relatively strong.
However, the nEPT may still provide larger flexibility than the EPT (requiring exact commutative local symmetry) in designing CLSs for systems with non-intraconnected symmetric subparts.

Comparing the nEPT with the EPT, some similarities but also subtle differences become evident. 
Both the nEPT and the EPT block-diagonalize the Hamiltonian, and in both cases the eigenstates are decomposed into two classes:
extended states generally occupying all sites of the system, and CLSs localized on $\SS$ (the sites of the symmetric subsystems, non-trivially affected by the permutation $\symmetryOp$). 
However, the detailed properties of eigenstates in each class are different for the EPT and nEPT.
Extended eigenstates (the $\bm{\phi}^{\nu \in [1,f+l]}$ in \cref{eq:eigenstateDecomposition,eq:eigenstateDecompositionSIAMSecondClass}) are symmetric under the action of $\symmetryOp$ for the EPT, while this holds only for equal $\gamma_{i}$ for the nEPT (in which case $\symmetryOp$ becomes commutative and the EPT applies).
Also, CLSs (the $\bm{\phi}^{f+ml+r}$ in \cref{eq:eigenstateDecomposition,eq:eigenstateDecompositionSIAMSecondClass}) determined by the nEPT are more compactly localized, on only a subset of $\SS$, as the $k-1-m$ vectors $\bm{0}_{l}$ in \cref{eq:eigenstateDecompositionSIAMSecondClass} indicate.

\section{Compact localized eigenstates in lattice systems} \label{sec:applications}

Having presented and analyzed the (n)EPT and its implications for the eigenspectra and eigenstates of discrete models with (restricted non)commutative local symmetries, in the following we demonstrate concrete applications to compact state design in extended lattice systems.

\subsection{Engineering bound states in the continuum} \label{sec:SymmetryAndDisorder}

The band structure of a periodic lattice provides energetic continua for extended (Bloch) eigenstates respecting the underlying discrete translational symmetry.
In this section we will demonstrate how certain perturbations, which destroy the periodic character of the lattice, may nevertheless leave the band structure of the system unchanged. 
Key to this are tailored local perturbations of one or more unit cells, which can be described by local symmetries and thereby induce the occurrence of CLSs.

\begin{figure}[t]
	\centering
	\includegraphics[max size={.95\linewidth}{0.5 \textheight}]{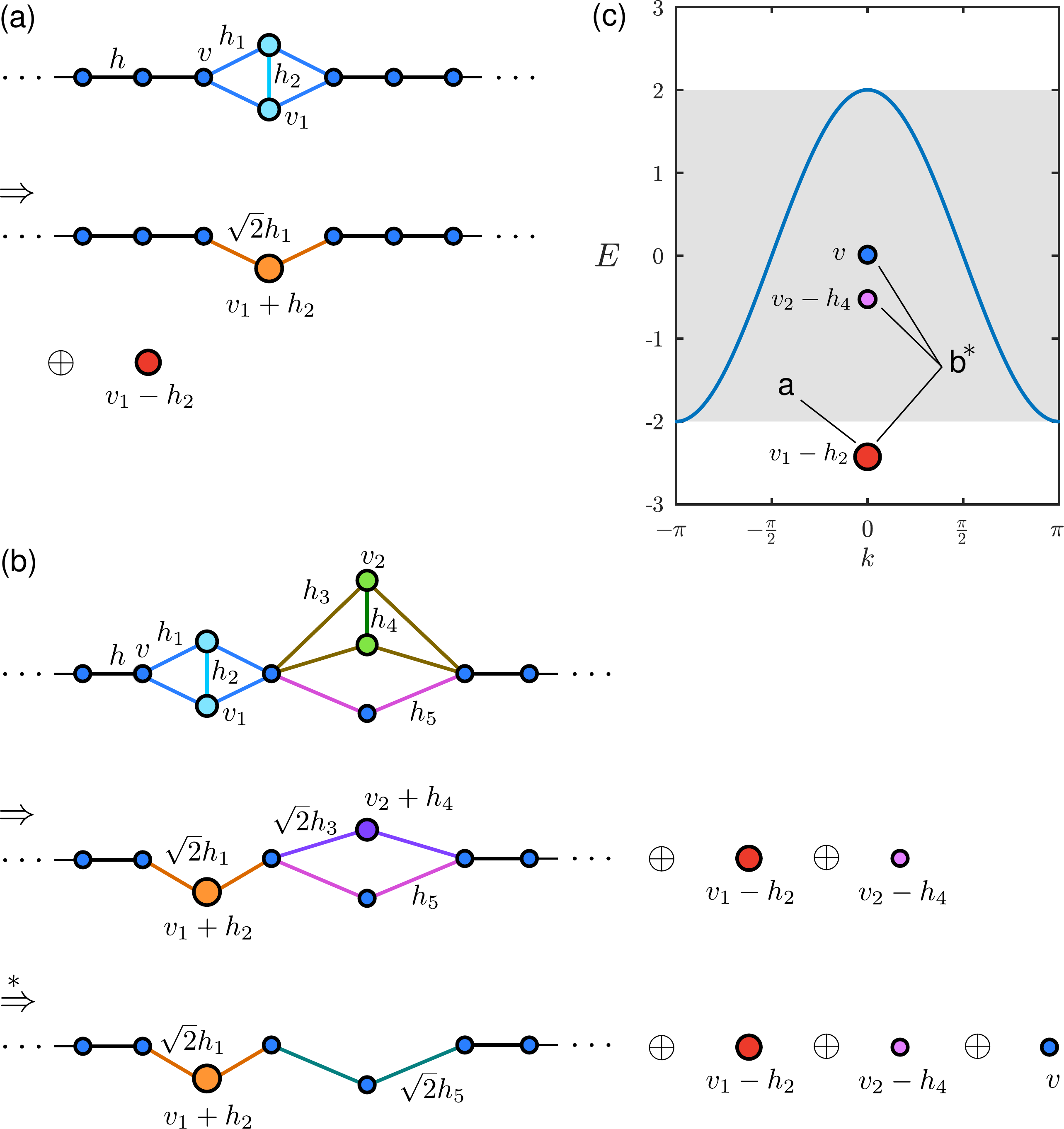}
	\caption{
	{\bf (a)} Periodic lattice system locally perturbed by a symmetric dimer defect with indicated onsite and hopping elements (top).
	The corresponding Hamiltonian can be transformed (bottom) via the EPT into the direct product of the divisor matrix $R$, corresponding to a generally perturbed linear chain, and the $1 \times 1$ matrix $B_1$ corresponding to a CLS on the defect with energy $E_{\text{CLS}} = v_{1} - h_{2}$.
	For the special case of $h_{1} = h/\sqrt{2}$ and $h_{2} = v - v_{1}$, the spectrum $\sigma(H)$ consists of the unperturbed chain band structure with an additional tunable bound state energy $E_{\text{CLS}}$. 
	{\bf (b)} The same chain with an additional, different, locally symmetric perturbation next to the first one, which can be treated by an iterative decomposition. 
	The first decomposition ($\Rightarrow$) reveals the occurrence of two CLEs at energies $v_{1} - h_{2}, v_{2} - h_{4}$. The second decomposition ($\xRightarrow{*}$) applies if $v_{2} + h_{4} = v, \; \sqrt{2} h_{3} = h_{5}$. 
	For parameters tuned so that $\sqrt{2} h_{1} = h, \; v_{1} + h_{2} = v, \; \sqrt{2} h_{5} = h$, the spectrum of the perturbed system again consists of the unperturbed chain band structure and three CLS energies $v_{1} - h_{2}, v_{2} -h_{4}, v$.
	{\bf (c)} Band structure of the unperturbed chain (blue line), present in the spectra of the tuned locally symmetric systems of (a) and (b), together with indicated corresponding CLS eigenenergies.
	Different onsite and hopping elements are depicted with different sizes and colors.
	}
	\label{fig:bic}
\end{figure}

Let us consider the system depicted in \cref{fig:bic}\,(a): 
A tight-binding periodic chain (with $v_n = v$ and $h_{n,n \pm 1} = h$) perturbed locally by replacing a lattice site with a dimer of onsite energy $v_{1}$ and intra-hopping $h_{2}$, in turn connected to the chain by hoppings $h_{1}$.
For generic defect parameters, the Bloch states of the unperturbed chain are no longer eigenstates of the system, and defect modes with exponential decay into the left and right semi-infinite chains arise. 
This may change, however, if the defect forms a commutative local symmetry, as we now demonstrate. 
In this case, the EPT provides a symmetry-adapted partitioning of the Hamiltonian into the matrices $R$ and $B_1$, as shown graphically in \cref{fig:bic}\,(a).
The divisor matrix $R$ corresponds to a linear chain with a single-site defect of energy $v_{1} + h_{2}$, connected by hoppings $\sqrt{2}h_{1}$, while $B_1$ corresponds to a single CLS, localized only on the dimer, with energy $E_{\text{CLS}} = v_{1}-h_{2}$, as indicated in \cref{fig:bic}\,(c) \cite{PythTB}. 
Following this partitioning, the spectrum is given by $\sigma(H) = \sigma(R) \cup \sigma(B_1)$.

As we see, also in the partitioned representation a defect is generally retained in the chain, which would lead to, e.\,g., backscattering of incident waves lying energetically in the unperturbed continuum.
Note, however, that the defect parameters can be tuned so as to effectively recover those of the unperturbed chain:
Setting $h_{1} = h/\sqrt{2}$ and $h_{2} = v - v_{1}$ indeed makes $R$ coincide with the unperturbed Hamiltonian.
Thus, despite the presence of the defect, the spectrum in this case consists of the band structure of the unperturbed chain, augmented by the energy of the CLS.
Moreover, by simultaneously tuning $v_{1}$ and $h_{2}$ such that $h_{2} = v - v_{1}$, the CLS can be moved in energy into the band of the chain, so that it becomes `bound state in the continuum' \cite{Vonneuman1929_PZ_30_467_UberMerkwuerdige,Stillinger1975_PRA_11_446_BoundStates}.
In the present case, this state does not interact with the extended continuum states due to eigenvector orthogonality, and thus the defect is effectively invisible for an incident wave (i.\,e. causes no backscattering).

Such a `renormalization' of defects into unperturbed chain sites was recently shown to explain the absence of localization \cite{Pal2013_EEL_102_17004_CompleteAbsence}, though for the special case of one dimension and zero intra-dimer coupling $h_{2}$.
Following the above paradigm, the EPT can be used to easily generalize the approach to perturbations of various complexity and connectivity as well as to higher dimensions.
The key for such a generalization is to have a perturbation that renders the divisor matrix $R$ identical to the unperturbed Hamiltonian. Note that this can be done even for different kinds of perturbations, as shown in \cref{fig:bic}\,(b). 
Here, an iterative decomposition is possible, provided that $v_{2} + h_{4} = v,\; \sqrt{2}h_{3} = h_{5}$. 
If, additionally, $\sqrt{2}h_{1} = h,\; v_{1} +h_{2} = v,\; \sqrt{2} h_{5} = h$ (indicated by an asterisk in \cref{fig:bic}\,(b)), then the original band structure of the unperturbed chain is recovered, together with three additional bound states at energies $v_{1} - h_{2}, v_{2} -h_{4}, v$, as shown in \cref{fig:bic}\,(c).

CLSs tailored as above to be `invisible' to a host lattice can clearly be inserted in multiple positions in the lattice without affecting the unperturbed band structure.
Notably, the same could be done for restricted noncommutative local symmetry defects using the nEPT, as long as its conditions are met. 
This concept of tailoring $R$ is hereby neither limited by the number of dimensions nor by the number of perturbed unit cells, and thus applies to quite generic extended lattice models.
As an application, CLSs could be distributed along a given aperiodic or even random sequence, to then study their interaction with continuum states by gradually breaking the local symmetry of the defects.

\subsection{Using symmetries to design flat bands} \label{sec:designFlatBands}

The above engineering of bound states in an unperturbed continuum via the (n)EPT relies on making the divisor matrix $R$ coincide with the unperturbed lattice Hamiltonian by tuning the defect parameters.
To obtain a band structure for generic defect parameters, however, the defects need to be placed periodically as well.
Then, since the corresponding CLSs vanish on the sites (fixed under the local symmetry $\symmetryOp$) connected to adjacent lattice cells, their energy will also be independent of the Bloch momentum.
Consequently, a flat band will form at each CLS eigenenergy. 
An example for this is the well-known one-dimensional diamond ladder lattice \cite{Flach2014_EPL_105_30001_DetanglingFlat} which can be constructed by periodically repeating the perturbed unit cell in \cref{fig:bic}\,(a) and which features global chiral symmetry \cite{Ramachandran2017_PRB_96_161104_Chiralflatbands};
similarly, cross-stitch and one-dimensional pyrochlore lattices \cite{Flach2014_EPL_105_30001_DetanglingFlat} can be treated with the present local symmetry approach.

\begin{figure}[t]
	\centering
	\includegraphics[max size={.95\linewidth}{0.7 \textheight}]{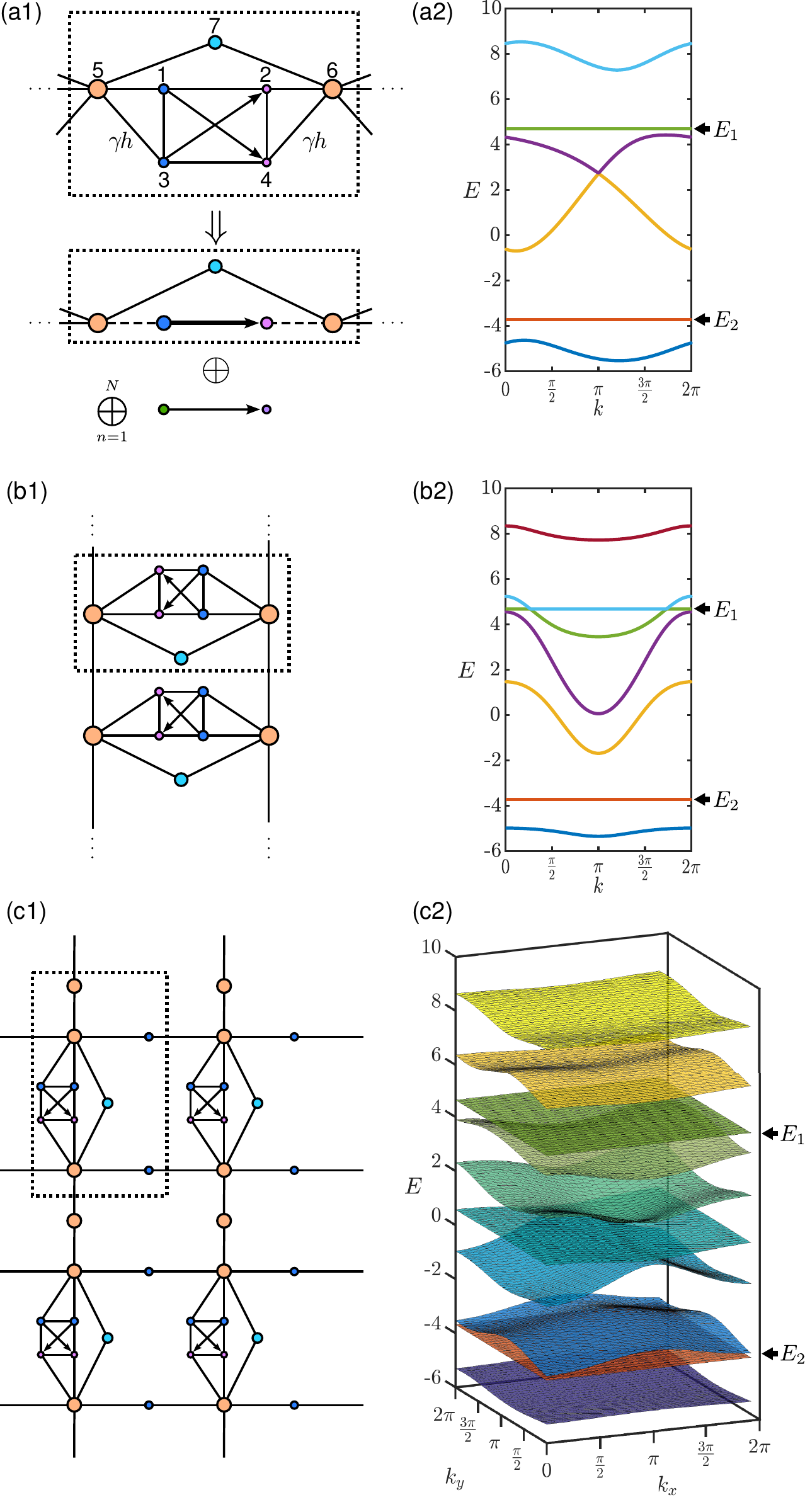}
	\caption{
	{\bf (a1)} Unit cell (indicated by dotted rectangle) of a periodic setup, with a local symmetry under the permutation $\symmetryOp$: $1 \leftrightarrow 3$, $2 \leftrightarrow 4$ leading to CLSs at two energies $E_1,E_2$, and
	{\bf (a2)} corresponding band structure with flat bands at $E_1,E_2$.
	{\bf (b1)--(b2)} Like in (a1)--(a2) but with the same unit cell connected in parallel via two sites to each neighboring cell.
	{\bf (c1)--(c2)} Like in (a1)--(a2) but with the same unit cell augmented by three sites and connected into a 2D lattice.
	While the dispersive bands are different in each case, the flat bands remain at $E_{1,2}$ since the symmetric subsystem remains unchanged. 
	In the lower part of (a1), the EPT decomposition ($\Rightarrow$) of the system's Hamiltonian into that of a modified lattice and ($\oplus$) $N \to \infty$ copies of isolated dimers (with eigenenergies $E_{1,2}$) is visualized. 
	Arrowed lines indicate complex-valued hoppings and dashed lines modified real hoppings. 
	For an asymmetry parameter $\gamma \neq 1$ (indicated in (a1)) and vanishing intraconnections ($h_{mn} = 0; m,n = 1,2,3,4$), the nEPT applies (see text).
	}
	\label{fig:FlatBands}
\end{figure}

We now show how the (n)EPT can be used to design lattices in arbitrary dimensions hosting a prescribed number of flat bands at desired energies. 
Consider a lattice like the one in \cref{fig:FlatBands}\,(a1), featuring a commutative local symmetry for $\gamma = 1$ in each unit cell under the permutation $\symmetryOp$: $1 \leftrightarrow 3$, $2 \leftrightarrow 4$ (leaving all other sites fixed). 
Then, by the EPT there are $(k-1)l = 2$ CLSs localized on the sites $\SS = \{1,2,3,4 \}$ within the unit cell, with $k$ and $l$ being the uniform size and number of non-trivial orbits of $\symmetryOp$, respectively. 
Two flat bands thus form at the CLS energies $E_{1,2}$, as shown in \cref{fig:FlatBands}\,(a2) \cite{PythTB}.
Specifically, $E_{1,2}$ are the eigenvalues of the matrix
\begin{equation} \label{eq:FBEigenvalues}
	B_{1} = 
	\begin{bmatrix}
	v_{1} & h_{1,2} \\
	h_{2,1} & v_{2}
	\end{bmatrix}
	-
	\begin{bmatrix}
	h_{1,3} & h_{1,4} \\
	h_{2,3} & h_{2,4}
	\end{bmatrix},
\end{equation}
of \cref{eq:DivisorMatrix}, and thus depend only on the elements within the subpart $\SS$ of the unit cell.
Note here that, for $H$ to be Hermitian and $\symmetryOp$-symmetric, $h_{1,3}$ and $h_{2,4}$ need to be real, while $h_{2,3} = h_{4,1}$ may as well be complex (indicated by arrows).

In complete analogy, flat bands form in the presence of restricted noncommutative local symmetries in the unit cell following the conditions of the nEPT.
Specifically, for $\gamma \neq 1$ in \cref{fig:FlatBands}\,(a1), but with vanishing local symmetry intraconnection ($h_{1,3},h_{2,4},h_{2,3},h_{4,1} = 0$), we obtain two flat bands (not shown) at the CLS eigenvalues of the matrix 
\begin{equation} 
	C_{0} = \begin{bmatrix}
	v_{1} & h_{1,2} \\
	h_{2,1} & v_{2}
	\end{bmatrix}
\end{equation}
defined in \cref{eq:DefinitionsOfSubmatrices}.
Note that those are \emph{independent} of the asymmetry factor $\gamma$. 

Figure \ref{fig:FlatBands}\,(b1) shows a lattice where the same unit cell as before is connected `in parallel' (via two connections) to each neighboring unit cell, instead of `serially' as in \cref{fig:FlatBands}\,(a1). 
Again, the unit cell's CLSs lead to two tunable flat bands at energies given by the eigenvalues of \cref{eq:FBEigenvalues}, as shown in \cref{fig:FlatBands}\,(b2).

In the two-dimensional (2D) example of \cref{fig:FlatBands}\,(c1), the unit cell differs from that of \cref{fig:FlatBands}\,(a1) in that it contains three additional sites, though still containing the same locally symmetric unit.
The CLS eigenenergies are also here independent of the Bloch momentum, now in both directions of translational invariance, and 2D corresponding bands thus arise. 
Note that their position is the same as in \cref{fig:FlatBands}\,(a) and (b), since the underlying symmetric substructure is not changed.

Concluding the above, we have shown that a lattice automatically features one or more symmetry-induced flat bands if (i) the unit cell possesses a commutative or restricted noncommutative local symmetry and (ii) this symmetry is unbroken when isolated unit cells are connected to form the lattice. 
Note that this approach to flat bands can be related to the common description of symmetry via the point group of the unit cell, whose action leaves at least one point fixed.
Indeed, the discreteness of the considered model Hamiltonian maps each point group element to a site permutation. 
This constitutes then a global symmetry of the isolated unit cell, and thus a special case of condition (i) above.
In turn, condition (ii) is fulfilled provided that the unit cells are connected through sites located at the point group's fixed points.
This link of the proposed approach to point groups may aid the description and design of symmetry-induced flat bands in more complex lattice systems.

Having seen how local symmetries lead to $(k-1)l$ flat bands ($k$ being the number of copies of $C_{0} \in \mathbb{C}^{l \times l}$ defined in \cref{eq:reorderingOfHamiltonian,eq:SIAMHamiltonian}), let us now look at the remaining bands. 
Note that these are usually completely dispersive, but could contain flat bands as well that are induced by other means than the above (restricted non)commutative local symmetries. 
In any case, all of these remaining bands are entirely determined by the divisor matrix $R$ which is explicitly represented graphically in \cref{fig:FlatBands}\,(a1) (between $\Downarrow$ and $\oplus$). 
Thus, these remaining bands can be directly obtained by diagonalizing the matrix $R$ which represents a strictly periodic system. 
Note that the partition into matrices $R$ and $B_{j}$ (by the EPT) or $C_{0}$ (by the nEPT) allows for an effective design process in which the symmetry-induced flat bands and the remaining band structure can be designed \emph{separately}. 

It is clear that the present approach to design flat bands using the EPT or nEPT applies to the class of lattices containing commutative or restricted noncommutative local symmetries. 
Thus, it does not cover other cases of lattices with other classes of local symmetries, which may also host flat bands generated by the very general method developed in Ref.\,\cite{Maimaiti2017_PRB_95_115135_CompactLocalized}.
The essence of the present approach is that, instead of generating flat bands from conditions imposed on the site-resolved eigenvalue problem (\ref{eq:evp}), it is based on unified and intuitive symmetry principles forcing the occurrence of CLSs.
We thus view it as a complementary method which lends an insightful understanding to already existing methods.

\section{Conclusions} \label{sec:conclusions}

We have shown how two very recent results from graph theory can be used to analyze discrete Hamiltonians with local symmetries.
The resulting framework demonstrates the impact of two types of local symmetries on the eigenstates of a Hamiltonian $H$, including the formation of so-called compact localized states (CLSs). 
These two types of local symmetries are described by site permutations which either leave all (commutative local symmetries) or some (restricted noncommutative local symmetries) matrix elements of $H$ invariant. 
More specifically, the restricted noncommutative local symmetries are such that the symmetric subsystems are (i) not interconnected and (ii) asymmetrically coupled to the remaining part of the system.

The essence of the framework is a symmetry-adapted partition of $H$ into smaller matrices $R$ and $B_{i}$ whose collective eigenvalue spectrum is equal to that of the original $H$. 
Depending on the exact character of the local symmetry, $H$ is assured to have one or more compact localized eigenstates which are localized on the symmetric subsystem $\SS$. 
Their energies are given by the spectra of the matrices $B_{i}$. 
All other eigenstates of $H$ are not localized on $\SS$, with their energy given by the spectrum of $R$.
In short, the framework provides the total eigenvalue spectrum as well as eigenvectors of the Hamiltonian in terms of symmetry-adapted submatrices, which are in turn more efficiently computed and better controllable by parametric tuning.

We apply this novel framework to tight binding systems and explicitly design flat bands at tailored energies in lattices of one and two dimensions, with the generalization to arbitrary dimensions being straightforward. 
Moreover, we use the methodology to demonstrate the occurrence of bound states in the energy continuum of a periodic chain perturbed by one or more symmetric defects. 
For both flat bands and bound states in the continuum, our results give an intuitive understanding of the impact of local symmetries, paired with a high degree of control over the respective energies.
We believe that the present framework may serve as complementary to existing methods in the design of CLSs and flat bands, by offering a unifying, intuitive, and efficient way to connect them to local symmetries.
\section{Acknowledgements}

Financial support by the Deutsche Forschungsgemeinschaft under grant DFG Schm 885/29-1 is gratefully acknowledged. 
M.R. gratefully acknowledges financial support by the `Stiftung der deutschen Wirtschaft' in the framework of a scholarship.
We thank C. Fey for discussions on the proof of the nonequitable partition theorem.


\appendix*

\section{Proof of the nEPT}  \label[appendix]{ProofAsymEPT}

We here prove property \ref{point:esDecompositionnEPT} of the nEPT.
To this end, we need to show that the vectors
\begin{equation} \label{eq:eigenstateDecompositionSIAMSecondClassProof}
\bm{x}_{j} =
\begin{bmatrix}[1.5]
\bm{w}_{j} \\
\frac{\gamma_{1}}{\sqrt{\xi}} \bm{v}_{j} \\
\frac{\gamma_{2}}{\sqrt{\xi}}\bm{v}_{j} \\
\vdots \\
\frac{\gamma_{k}}{\sqrt{\xi}}\bm{v}_{j}
\end{bmatrix},~
\bm{y}_{m,r} =
\begin{bmatrix}[1.5]
\bm{0}_{f} \\
\frac{\gamma_{1}}{\gamma_{1}} \bm{u}_{0,r} \\
\frac{\gamma_{2}}{\gamma_{1}} \bm{u}_{0,r}\\
\vdots \\
\frac{\gamma_{m}}{\gamma_{1}} \bm{u}_{0,r}\\
-\frac{\sum_{i=1}^{m} \gamma_{i}^{2} }{\gamma_{1} \gamma_{m+1}} \bm{u}_{0,r}\\
\bm{0}_{l} \\
\vdots \\
\bm{0}_{l} 
\end{bmatrix},
\end{equation}
with $j \in [1,f+l], m \in [1,k-1], r \in [1,l]$, are linearly independent eigenvectors of the Hamiltonian $H'$ given in \cref{eq:HBlockDiagonalSIAM}.
 
Since ${\footnotesize \begin{bmatrix}[.8] \bm{w}_j \\[-.5ex] \bm{v}_j \end{bmatrix}}$ 
is the $j$-th eigenvector of the divisor matrix $R$ (given in \cref{eq:HBlockDiagonalSIAM}) with eigenvalue $\lambda_{j}$, i.\,e.
\begin{equation}
  \begin{bmatrix}
  F & \sqrt{\xi} \cdot G\\
  \sqrt{\xi} \cdot G^{\dagger} & C_{0}
  \end{bmatrix} \begin{bmatrix}
  \bm{w}_{j} \\
  \bm{v}_{j}
  \end{bmatrix}
  = 
  \begin{bmatrix}
  F \bm{w}_{j} + \sqrt{\xi} \cdot G \bm{v}_{j} \\
  \sqrt{\xi} \cdot G^{\dagger} \bm{w}_{j} + C_{0} \bm{v}_{j}
  \end{bmatrix}
  = \lambda_{j} \begin{bmatrix}
  \bm{w}_{j} \\
  \bm{v}_{j}
  \end{bmatrix}, \nonumber
\end{equation}
applying $H'$ on $\bm{x}_{j}$ yields
\begin{align}
      H' \bm{x}_{j} 
      = H \begin{bmatrix}[1.4]
      \bm{w}_{j} \\
      \frac{\gamma_{1}}{\sqrt{\xi}} \bm{v}_{j} \\
      \vdots \\
      \frac{\gamma_{k}}{\sqrt{\xi}}\bm{v}_{j}
      \end{bmatrix}
      = \begin{bmatrix}[1.5]
      F \bm{w}_{j} + \frac{\sum_{i=1}^{k} |\gamma_{i}|^{2}}{ \sqrt{\xi}} G \bm{v}_{j} \\
      \gamma_{1} G^{\dagger} \bm{w}_{j} + \frac{\gamma_{1}}{\sqrt{\xi}} C_{0} \bm{v}_{j} \\
      \vdots \\
      \gamma_{k} G^{\dagger} \bm{w}_{j} + \frac{\gamma_{k}}{ \sqrt{\xi}} C_{0} \bm{v}_{j}
      \end{bmatrix} \nonumber \\
      = \begin{bmatrix}[1.5]
      F \bm{w}_{j} + \sqrt{\xi} \cdot G \bm{v}_{j} \\
      \gamma_{1} G^{\dagger} \bm{w}_{j} + \frac{\gamma_{1}}{\sqrt{\xi}} C_{0} \bm{v}_{j} \\
      \vdots \\
      \gamma_{k} G^{\dagger} \bm{w}_{j} + \frac{\gamma_{k}}{ \sqrt{\xi}} C_{0} \bm{v}_{j}
      \end{bmatrix}
      =\lambda_{j} \begin{bmatrix}[1.4]
      \bm{w}_{j} \\
      \frac{\gamma_{1}}{\sqrt{\xi}} \bm{v}_{j} \\
      \vdots \\
      \frac{\gamma_{k}}{\sqrt{\xi}}\bm{v}_{j}
      \end{bmatrix}
      = \lambda_{j} \bm{x}_{j} .
\end{align}
Thus, $\{\bm{x}_{j}\}$ are eigenvectors of $H'$. 
If we choose the set of eigenvectors 
$\left\{ {\footnotesize \begin{bmatrix}[.8] \bm{w}_j \\[-.5ex] \bm{v}_j \end{bmatrix}} \right\}$ 
such that they are pairwise linearly independent (which can always be done), then this is also the case for the set $\{ \bm{x}_{j}\}$. To see this, let us assume that there exists an $\bm{x}_{i}$ and constants $\{\alpha_{j}\}$ such that $\bm{x}_{i}$ is given by a superposition of $\{\bm{x}_{j}\}$ with $j \ne i$, i.\,e.
\begin{equation}
\bm{x}_{i} = \sum_{j \ne i} \alpha_{j} \bm{x}_{j} .
\end{equation}
Then, from the definition of $\bm{x}_{i}$, it would hold that
\begin{equation}
\begin{bmatrix}
\bm{w}_{i} \\
\frac{\gamma_{1}}{\sqrt{\xi}}\bm{v}_{i}
\end{bmatrix} =
\sum_{j\ne i} \alpha_{j} \begin{bmatrix}
\bm{w}_{j} \\
\frac{\gamma_{1}}{\sqrt{\xi}} \bm{v}_{j}
\end{bmatrix} \Rightarrow
\begin{bmatrix}
\bm{w}_{i} \\
\bm{v}_{i}
\end{bmatrix} =
\sum_{j\ne i} \alpha_{j} \begin{bmatrix}
\bm{w}_{j} \\
\bm{v}_{j}
\end{bmatrix},
\end{equation}
which is not true since the 
${\footnotesize \begin{bmatrix}[.8] \bm{w}_j \\[-.5ex] \bm{v}_j \end{bmatrix}}$ 
are pairwise linearly independent. 
Thus, the $\bm{x}_{j}$ are pairwise linearly independent eigenvectors of $H'$.

Further, since $C_{0} \bm{u}_{0,r} = \lambda_{0,r} \bm{u}_{0,r}$ by definition, application of $H'$ on $\bm{y}_{m,r}$ yields \\
\begin{equation}
H'
\begin{bmatrix}
\bm{0}_f \\
a_{1} \bm{u}_{0,r} \\
\vdots \\
a_{k} \bm{u}_{0,r}
\end{bmatrix}
=
\begin{bmatrix}
C_{0}\bm{0}_f \\
a_{1} C_{0}\bm{u}_{0,r} \\
\vdots \\
a_{k} C_{0} \bm{u}_{0,r}
\end{bmatrix} =
\lambda_{0,r} 	\begin{bmatrix}
\bm{0}_f \\
a_{1} \bm{u}_{0,r} \\
\vdots \\
a_{k} \bm{u}_{0,r}
\end{bmatrix}
\end{equation}
for any $a_{1},\ldots, a_{k} \in \mathbb{C}$. 
Thus, the $ \bm{y}_{m,r}$ are eigenvectors of $H'$. 
As easily shown, they are also pairwise orthogonal, both for the same and for different $\bm{u}_{0,r}$.

Having shown that both sets $\{\bm{y}_{m,r}\},\{\bm{x}_{j}\}$ are eigenvectors of $H'$, we need to show that they form a linearly independent set. 
Evaluating the scalar product of $\bm{x}_{j}$ and $\bm{y}_{m=i,r}$ for an arbitrary $i$ and $r$ gives
\begin{equation}
\bm{x}_{j} \cdot \bm{y}_{i,r}
= \left( \frac{\sum_{p=1}^{i} \gamma_{p}^{2} }{\gamma_{1} \sqrt{\xi}} - \frac{\gamma_{i+1}}{\sqrt{\xi}}\frac{\sum_{p=1}^{i} \gamma_{p}^{2} }{\gamma_{1} \gamma_{i+1}} \right)\bm{v}_{j} \cdot \bm{u}_{0,r}  = 0,
\end{equation}
where the last equality stems from the cancellation of the two summands in parentheses.
Note that $H' \in \mathbb{C}^{N\times N}$ has $N = f + kl$ linearly independent eigenstates. Since $R\in \mathbb{C}^{n+l}$ and since $\{\bm{y}_{m,r}\}$ contains $(k-1)l$ orthogonal eigenstates, we have thus proven that the eigenstates of $H'$ are given by $\{\bm{x}_{j}\},\{\bm{y}_{m,r}\}$.

\vfill




\begin{thebibliography}{52}
\expandafter\ifx\csname natexlab\endcsname\relax\def\natexlab#1{#1}\fi
\expandafter\ifx\csname bibnamefont\endcsname\relax
  \def\bibnamefont#1{#1}\fi
\expandafter\ifx\csname bibfnamefont\endcsname\relax
  \def\bibfnamefont#1{#1}\fi
\expandafter\ifx\csname citenamefont\endcsname\relax
  \def\citenamefont#1{#1}\fi
\expandafter\ifx\csname url\endcsname\relax
  \def\url#1{\texttt{#1}}\fi
\expandafter\ifx\csname urlprefix\endcsname\relax\def\urlprefix{URL }\fi
\providecommand{\bibinfo}[2]{#2}
\providecommand{\eprint}[2][]{\url{#2}}

\bibitem[{\citenamefont{Flach et~al.}(2014)\citenamefont{Flach, Leykam,
  Bodyfelt, Matthies, and
  Desyatnikov}}]{Flach2014_EPL_105_30001_DetanglingFlat}
\bibinfo{author}{\bibfnamefont{S.}~\bibnamefont{Flach}},
  \bibinfo{author}{\bibfnamefont{D.}~\bibnamefont{Leykam}},
  \bibinfo{author}{\bibfnamefont{J.~D.} \bibnamefont{Bodyfelt}},
  \bibinfo{author}{\bibfnamefont{P.}~\bibnamefont{Matthies}}, \bibnamefont{and}
  \bibinfo{author}{\bibfnamefont{A.~S.} \bibnamefont{Desyatnikov}},
  \bibinfo{journal}{Europhys. Lett.} \textbf{\bibinfo{volume}{105}},
  \bibinfo{pages}{30001} (\bibinfo{year}{2014}).

\bibitem[{\citenamefont{Maimaiti et~al.}(2017)\citenamefont{Maimaiti,
  Andreanov, Park, Gendelman, and
  Flach}}]{Maimaiti2017_PRB_95_115135_CompactLocalized}
\bibinfo{author}{\bibfnamefont{W.}~\bibnamefont{Maimaiti}},
  \bibinfo{author}{\bibfnamefont{A.}~\bibnamefont{Andreanov}},
  \bibinfo{author}{\bibfnamefont{H.~C.} \bibnamefont{Park}},
  \bibinfo{author}{\bibfnamefont{O.}~\bibnamefont{Gendelman}},
  \bibnamefont{and} \bibinfo{author}{\bibfnamefont{S.}~\bibnamefont{Flach}},
  \bibinfo{journal}{Phys. Rev. B} \textbf{\bibinfo{volume}{95}},
  \bibinfo{pages}{115135} (\bibinfo{year}{2017}).

\bibitem[{\citenamefont{Anderson}(1958)}]{Anderson1958_PR_109_1492_AbsenceDiffusion}
\bibinfo{author}{\bibfnamefont{P.~W.} \bibnamefont{Anderson}},
  \bibinfo{journal}{Phys. Rev.} \textbf{\bibinfo{volume}{109}},
  \bibinfo{pages}{1492} (\bibinfo{year}{1958}).

\bibitem[{\citenamefont{Sutherland}(1986)}]{Sutherland1986_PRB_34_5208-5211_Localizationelectronicwave}
\bibinfo{author}{\bibfnamefont{B.}~\bibnamefont{Sutherland}},
\bibinfo{journal}{Phys. Rev. B} \textbf{\bibinfo{volume}{34}},
\bibinfo{pages}{5208} (\bibinfo{year}{1986}).

\bibitem[{\citenamefont{Bergman et~al.}(2008)\citenamefont{Bergman, Wu, and
  Balents}}]{Bergman2008_PRB_78_125104_BandTouching}
\bibinfo{author}{\bibfnamefont{D.~L.} \bibnamefont{Bergman}},
  \bibinfo{author}{\bibfnamefont{C.}~\bibnamefont{Wu}}, \bibnamefont{and}
  \bibinfo{author}{\bibfnamefont{L.}~\bibnamefont{Balents}},
  \bibinfo{journal}{Phys. Rev. B} \textbf{\bibinfo{volume}{78}},
  \bibinfo{pages}{125104} (\bibinfo{year}{2008}).

\bibitem[{\citenamefont{Derzhko and
  Richter}(2006)}]{Derzhko2006_EPJB-_52_23_UniversalLow-temperature}
\bibinfo{author}{\bibfnamefont{O.}~\bibnamefont{Derzhko}} \bibnamefont{and}
  \bibinfo{author}{\bibfnamefont{J.}~\bibnamefont{Richter}},
  \bibinfo{journal}{Eur. Phys. J. B} \textbf{\bibinfo{volume}{52}},
  \bibinfo{pages}{23} (\bibinfo{year}{2006}).

\bibitem[{\citenamefont{Derzhko et~al.}(2010)\citenamefont{Derzhko, Richter,
  Honecker, Maksymenko, and
  Moessner}}]{Derzhko2010_PRB_81_014421_Low-temperatureProperties}
\bibinfo{author}{\bibfnamefont{O.}~\bibnamefont{Derzhko}},
  \bibinfo{author}{\bibfnamefont{J.}~\bibnamefont{Richter}},
  \bibinfo{author}{\bibfnamefont{A.}~\bibnamefont{Honecker}},
  \bibinfo{author}{\bibfnamefont{M.}~\bibnamefont{Maksymenko}},
  \bibnamefont{and} \bibinfo{author}{\bibfnamefont{R.}~\bibnamefont{Moessner}},
  \bibinfo{journal}{Phys. Rev. B} \textbf{\bibinfo{volume}{81}},
  \bibinfo{pages}{014421} (\bibinfo{year}{2010}).

\bibitem[{\citenamefont{Vicencio and
  {Mej\'{i}a-Cort\'{e}s}}(2014)}]{Vicencio2014_JO_16_015706_Diffraction-freeImage}
\bibinfo{author}{\bibfnamefont{R.~A.} \bibnamefont{Vicencio}} \bibnamefont{and}
  \bibinfo{author}{\bibfnamefont{C.}~\bibnamefont{{Mej\'{i}a-Cort\'{e}s}}},
  \bibinfo{journal}{J. Opt.} \textbf{\bibinfo{volume}{16}},
  \bibinfo{pages}{015706} (\bibinfo{year}{2014}).

\bibitem[{\citenamefont{Rojas-Rojas et~al.}(2017)\citenamefont{Rojas-Rojas, Morales-Inostroza, Vicencio, and Delgado}}]{Rojas-Rojas2017_PRA_96_043803_Quantumlocalizedstates}
\bibinfo{author}{\bibfnamefont{S.}~\bibnamefont{Rojas-Rojas}},
\bibinfo{author}{\bibfnamefont{L.}~\bibnamefont{Morales-Inostroza}},
\bibinfo{author}{\bibfnamefont{R.~A.} \bibnamefont{Vicencio}},
\bibnamefont{and} \bibinfo{author}{\bibfnamefont{A.}~\bibnamefont{Delgado}},
\bibinfo{journal}{Phys. Rev. A} \textbf{\bibinfo{volume}{96}},
\bibinfo{pages}{043803} (\bibinfo{year}{2017}).

\bibitem[{\citenamefont{Xia et~al.}(2016)\citenamefont{Xia, Hu, Song, Zong,
  Tang, and Chen}}]{Xia2016_OL_41_1435_DemonstrationFlat-band}
\bibinfo{author}{\bibfnamefont{S.}~\bibnamefont{Xia}},
  \bibinfo{author}{\bibfnamefont{Y.}~\bibnamefont{Hu}},
  \bibinfo{author}{\bibfnamefont{D.}~\bibnamefont{Song}},
  \bibinfo{author}{\bibfnamefont{Y.}~\bibnamefont{Zong}},
  \bibinfo{author}{\bibfnamefont{L.}~\bibnamefont{Tang}}, \bibnamefont{and}
  \bibinfo{author}{\bibfnamefont{Z.}~\bibnamefont{Chen}},
  \bibinfo{journal}{Opt. Lett.} \textbf{\bibinfo{volume}{41}},
  \bibinfo{pages}{1435} (\bibinfo{year}{2016}).

\bibitem[{\citenamefont{Vicencio et~al.}(2015)\citenamefont{Vicencio,
  Cantillano, {Morales-Inostroza}, Real, {Mej\'{i}a-Cort\'{e}s}, Weimann,
  Szameit, and Molina}}]{Vicencio2015_PRL_114_245503_ObservationLocalized}
\bibinfo{author}{\bibfnamefont{R.~A.} \bibnamefont{Vicencio}},
  \bibinfo{author}{\bibfnamefont{C.}~\bibnamefont{Cantillano}},
  \bibinfo{author}{\bibfnamefont{L.}~\bibnamefont{{Morales-Inostroza}}},
  \bibinfo{author}{\bibfnamefont{B.}~\bibnamefont{Real}},
  \bibinfo{author}{\bibfnamefont{C.}~\bibnamefont{{Mej\'{i}a-Cort\'{e}s}}},
  \bibinfo{author}{\bibfnamefont{S.}~\bibnamefont{Weimann}},
  \bibinfo{author}{\bibfnamefont{A.}~\bibnamefont{Szameit}}, \bibnamefont{and}
  \bibinfo{author}{\bibfnamefont{M.~I.} \bibnamefont{Molina}},
  \bibinfo{journal}{Phys. Rev. Lett.} \textbf{\bibinfo{volume}{114}},
  \bibinfo{pages}{245503} (\bibinfo{year}{2015}).

\bibitem[{\citenamefont{von Neuman and
  Wigner}(1929)}]{Vonneuman1929_PZ_30_467_UberMerkwuerdige}
\bibinfo{author}{\bibfnamefont{J.}~\bibnamefont{von Neuman}} \bibnamefont{and}
  \bibinfo{author}{\bibfnamefont{E.}~\bibnamefont{Wigner}},
  \bibinfo{journal}{Phys. Z.} \textbf{\bibinfo{volume}{30}},
  \bibinfo{pages}{465} (\bibinfo{year}{1929}).

\bibitem[{\citenamefont{Stillinger and
  Herrick}(1975)}]{Stillinger1975_PRA_11_446_BoundStates}
\bibinfo{author}{\bibfnamefont{F.~H.} \bibnamefont{Stillinger}}
  \bibnamefont{and} \bibinfo{author}{\bibfnamefont{D.~R.}
  \bibnamefont{Herrick}}, \bibinfo{journal}{Phys. Rev. A}
  \textbf{\bibinfo{volume}{11}}, \bibinfo{pages}{446} (\bibinfo{year}{1975}).
  
\bibitem[{\citenamefont{Hsu et~al.}(2017)\citenamefont{Hsu}}]{Hsu2016_NRM_1_16048_Boundstatescontinuum}
\bibinfo{author}{C. W. Hsu, B. Zhen, A. D. Stone, J. D. Joannopoulos, and M. Solja{\v c}i{\'c}},
\bibinfo{journal}{Nat. Rev. Mater.} \textbf{\bibinfo{volume}{1}},
\bibinfo{pages}{16048} (\bibinfo{year}{2016}).
 
  
\bibitem[{\citenamefont{Xiao et~al.}(2017)\citenamefont{Xiao, Ma, Zhang, and Chan}}]{Xiao2017_PRL_118_166803_TopologicalSubspaceInducedBound}
\bibinfo{author}{\bibfnamefont{Y.-X.} \bibnamefont{Xiao}},
\bibinfo{author}{\bibfnamefont{G.}~\bibnamefont{Ma}},
\bibinfo{author}{\bibfnamefont{Z.-Q.} \bibnamefont{Zhang}}, \bibnamefont{and}
\bibinfo{author}{\bibfnamefont{C.~T.} \bibnamefont{Chan}},
\bibinfo{journal}{Phys. Rev. Lett.} \textbf{\bibinfo{volume}{118}},
\bibinfo{pages}{166803} (\bibinfo{year}{2017}).

\bibitem[{\citenamefont{Klein and
  Mandal}(2015)}]{Klein2015_CMCC_74_247_LocalSymmetries}
\bibinfo{author}{\bibfnamefont{D.~J.} \bibnamefont{Klein}} \bibnamefont{and}
  \bibinfo{author}{\bibfnamefont{B.}~\bibnamefont{Mandal}},
  \bibinfo{journal}{MATCH Commun. Math. Comput. Chem.} \textbf{\bibinfo{volume}{74}},
  \bibinfo{pages}{247} (\bibinfo{year}{2015}).

\bibitem[{\citenamefont{Bodyfelt et~al.}(2014)\citenamefont{Bodyfelt, Leykam,
  Danieli, Yu, and Flach}}]{Bodyfelt2014_PRL_113_236403_FlatbandsUnder}
\bibinfo{author}{\bibfnamefont{J.~D.} \bibnamefont{Bodyfelt}},
  \bibinfo{author}{\bibfnamefont{D.}~\bibnamefont{Leykam}},
  \bibinfo{author}{\bibfnamefont{C.}~\bibnamefont{Danieli}},
  \bibinfo{author}{\bibfnamefont{X.}~\bibnamefont{Yu}}, \bibnamefont{and}
  \bibinfo{author}{\bibfnamefont{S.}~\bibnamefont{Flach}},
  \bibinfo{journal}{Phys. Rev. Lett.} \textbf{\bibinfo{volume}{113}},
  \bibinfo{pages}{236403} (\bibinfo{year}{2014}).

\bibitem[{\citenamefont{Parameswaran et~al.}(2013)\citenamefont{Parameswaran,
  Roy, and Sondhi}}]{Parameswaran2013_CRP_14_816_FractionalQuantum}
\bibinfo{author}{\bibfnamefont{S.~A.} \bibnamefont{Parameswaran}},
  \bibinfo{author}{\bibfnamefont{R.}~\bibnamefont{Roy}}, \bibnamefont{and}
  \bibinfo{author}{\bibfnamefont{S.~L.} \bibnamefont{Sondhi}},
  \bibinfo{journal}{C. R. Phys.} \textbf{\bibinfo{volume}{14}},
  \bibinfo{pages}{816} (\bibinfo{year}{2013}).

\bibitem[{\citenamefont{Yang et~al.}(2012)\citenamefont{Yang, Gu, Sun, and
  Das~Sarma}}]{Yang2012_PRB_86_241112_TopologicalFlat}
\bibinfo{author}{\bibfnamefont{S.}~\bibnamefont{Yang}},
  \bibinfo{author}{\bibfnamefont{Z.~C.}~\bibnamefont{Gu}},
  \bibinfo{author}{\bibfnamefont{K.}~\bibnamefont{Sun}}, \bibnamefont{and}
  \bibinfo{author}{\bibfnamefont{S.}~\bibnamefont{Das~Sarma}},
  \bibinfo{journal}{Phys. Rev. B} \textbf{\bibinfo{volume}{86}},
  \bibinfo{pages}{241112} (\bibinfo{year}{2012}).

\bibitem[{\citenamefont{Neupert et~al.}(2011)\citenamefont{Neupert, Santos,
  Chamon, and Mudry}}]{Neupert2011_PRL_106_236804_FractionalQuantum}
\bibinfo{author}{\bibfnamefont{T.}~\bibnamefont{Neupert}},
  \bibinfo{author}{\bibfnamefont{L.}~\bibnamefont{Santos}},
  \bibinfo{author}{\bibfnamefont{C.}~\bibnamefont{Chamon}}, \bibnamefont{and}
  \bibinfo{author}{\bibfnamefont{C.}~\bibnamefont{Mudry}},
  \bibinfo{journal}{Phys. Rev. Lett.} \textbf{\bibinfo{volume}{106}},
  \bibinfo{pages}{236804} (\bibinfo{year}{2011}).

\bibitem[{\citenamefont{Tang et~al.}(2011)\citenamefont{Tang, Mei, and
  Wen}}]{Tang2011_PRL_106_236802_High-temperatureFractional}
\bibinfo{author}{\bibfnamefont{E.}~\bibnamefont{Tang}},
  \bibinfo{author}{\bibfnamefont{J.~W.}~\bibnamefont{Mei}}, \bibnamefont{and}
  \bibinfo{author}{\bibfnamefont{X.~G.}~\bibnamefont{Wen}},
  \bibinfo{journal}{Phys. Rev. Lett.} \textbf{\bibinfo{volume}{106}},
  \bibinfo{pages}{236802} (\bibinfo{year}{2011}).

\bibitem[{\citenamefont{Souza and
  Herrmann}(2009)}]{Souza2009_PRB_79_153104_Flat-bandLocalization}
\bibinfo{author}{\bibfnamefont{A.~M.~C.} \bibnamefont{Souza}} \bibnamefont{and}
  \bibinfo{author}{\bibfnamefont{H.~J.} \bibnamefont{Herrmann}},
  \bibinfo{journal}{Phys. Rev. B} \textbf{\bibinfo{volume}{79}},
  \bibinfo{pages}{153104} (\bibinfo{year}{2009}).

\bibitem[{\citenamefont{Danieli et~al.}(2015)\citenamefont{Danieli, Bodyfelt,
  and Flach}}]{Danieli2015_PRB_91_235134_Flat-bandEngineering}
\bibinfo{author}{\bibfnamefont{C.}~\bibnamefont{Danieli}},
  \bibinfo{author}{\bibfnamefont{J.~D.} \bibnamefont{Bodyfelt}},
  \bibnamefont{and} \bibinfo{author}{\bibfnamefont{S.}~\bibnamefont{Flach}},
  \bibinfo{journal}{Phys. Rev. B} \textbf{\bibinfo{volume}{91}},
  \bibinfo{pages}{235134} (\bibinfo{year}{2015}).

\bibitem[{\citenamefont{Leykam et~al.}(2017)\citenamefont{Leykam, Flach, and
  Chong}}]{Leykam2017_PRB_96_064305_FlatBands}
\bibinfo{author}{\bibfnamefont{D.}~\bibnamefont{Leykam}},
  \bibinfo{author}{\bibfnamefont{S.}~\bibnamefont{Flach}}, \bibnamefont{and}
  \bibinfo{author}{\bibfnamefont{Y.~D.} \bibnamefont{Chong}},
  \bibinfo{journal}{Phys. Rev. B} \textbf{\bibinfo{volume}{96}},
  \bibinfo{pages}{064305} (\bibinfo{year}{2017}).

\bibitem[{\citenamefont{Ramezani}(2017)}]{Ramezani2017_PRA_96_011802_Non-hermiticity-induFlat}
\bibinfo{author}{\bibfnamefont{H.}~\bibnamefont{Ramezani}},
  \bibinfo{journal}{Phys. Rev. A} \textbf{\bibinfo{volume}{96}},
  \bibinfo{pages}{011802} (\bibinfo{year}{2017}).

\bibitem[{\citenamefont{Dias and
  Gouveia}(2015)}]{Dias2015_SR_5_srep16852_OrigamiRules}
\bibinfo{author}{\bibfnamefont{R.~G.} \bibnamefont{Dias}} \bibnamefont{and}
  \bibinfo{author}{\bibfnamefont{J.~D.} \bibnamefont{Gouveia}},
  \bibinfo{journal}{Sci. Rep.} \textbf{\bibinfo{volume}{5}},
  \bibinfo{pages}{16852} (\bibinfo{year}{2015}).

\bibitem[{\citenamefont{{Morales-Inostroza} and
  Vicencio}(2016)}]{Morales-inostroza2016_PRA_94_043831_SimpleMethod}
\bibinfo{author}{\bibfnamefont{L.}~\bibnamefont{{Morales-Inostroza}}}
  \bibnamefont{and} \bibinfo{author}{\bibfnamefont{R.~A.}
  \bibnamefont{Vicencio}}, \bibinfo{journal}{Phys. Rev. A}
  \textbf{\bibinfo{volume}{94}}, \bibinfo{pages}{043831}
  (\bibinfo{year}{2016}).

\bibitem[{\citenamefont{Ramachandran et~al.}(2017)\citenamefont{Ramachandran, Andreanov, and Flach}}]{Ramachandran2017_PRB_96_161104_Chiralflatbands}
\bibinfo{author}{\bibfnamefont{A.}~\bibnamefont{Ramachandran}},
\bibinfo{author}{\bibfnamefont{A.}~\bibnamefont{Andreanov}},
\bibnamefont{and} \bibinfo{author}{\bibfnamefont{S.}~\bibnamefont{Flach}},
\bibinfo{journal}{Phys. Rev. B} \textbf{\bibinfo{volume}{96}},
\bibinfo{pages}{161104} (\bibinfo{year}{2017}).

\bibitem[{\citenamefont{Xu et~al.}(2015)\citenamefont{Xu, Wang, Hang, Luo,
  Chan, and Lai}}]{Xu2015_SR_5_srep18181_DesignFull-k-space}
\bibinfo{author}{\bibfnamefont{C.}~\bibnamefont{Xu}},
  \bibinfo{author}{\bibfnamefont{G.}~\bibnamefont{Wang}},
  \bibinfo{author}{\bibfnamefont{Z.~H.} \bibnamefont{Hang}},
  \bibinfo{author}{\bibfnamefont{J.}~\bibnamefont{Luo}},
  \bibinfo{author}{\bibfnamefont{C.~T.} \bibnamefont{Chan}}, \bibnamefont{and}
  \bibinfo{author}{\bibfnamefont{Y.}~\bibnamefont{Lai}}, \bibinfo{journal}{Sci.
  Rep.} \textbf{\bibinfo{volume}{5}}, \bibinfo{pages}{18181}
  (\bibinfo{year}{2015}).

\bibitem[{\citenamefont{Kalozoumis
  et~al.}(2013{\natexlab{a}})\citenamefont{Kalozoumis, Morfonios, Diakonos, and
  Schmelcher}}]{Kalozoumis2013_PRA_87_032113_LocalSymmetries}
\bibinfo{author}{\bibfnamefont{P.~A.} \bibnamefont{Kalozoumis}},
  \bibinfo{author}{\bibfnamefont{C.}~\bibnamefont{Morfonios}},
  \bibinfo{author}{\bibfnamefont{F.~K.} \bibnamefont{Diakonos}},
  \bibnamefont{and}
  \bibinfo{author}{\bibfnamefont{P.}~\bibnamefont{Schmelcher}},
  \bibinfo{journal}{Phys. Rev. A} \textbf{\bibinfo{volume}{87}},
  \bibinfo{pages}{032113} (\bibinfo{year}{2013}{\natexlab{a}}).

\bibitem[{\citenamefont{Kalozoumis
  et~al.}(2013{\natexlab{b}})\citenamefont{Kalozoumis, Morfonios,
  Palaiodimopoulos, Diakonos, and
  Schmelcher}}]{Kalozoumis2013_PRA_88_033857_LocalSymmetries}
\bibinfo{author}{\bibfnamefont{P.~A.} \bibnamefont{Kalozoumis}},
  \bibinfo{author}{\bibfnamefont{C.}~\bibnamefont{Morfonios}},
  \bibinfo{author}{\bibfnamefont{N.}~\bibnamefont{Palaiodimopoulos}},
  \bibinfo{author}{\bibfnamefont{F.~K.} \bibnamefont{Diakonos}},
  \bibnamefont{and}
  \bibinfo{author}{\bibfnamefont{P.}~\bibnamefont{Schmelcher}},
  \bibinfo{journal}{Phys. Rev. A} \textbf{\bibinfo{volume}{88}},
  \bibinfo{pages}{033857} (\bibinfo{year}{2013}{\natexlab{b}}).

\bibitem[{\citenamefont{Kalozoumis et~al.}(2014)\citenamefont{Kalozoumis,
  Morfonios, Diakonos, and
  Schmelcher}}]{Kalozoumis2014_PRL_113_050403_InvariantsBroken}
\bibinfo{author}{\bibfnamefont{P.~A.} \bibnamefont{Kalozoumis}},
  \bibinfo{author}{\bibfnamefont{C.}~\bibnamefont{Morfonios}},
  \bibinfo{author}{\bibfnamefont{F.~K.} \bibnamefont{Diakonos}},
  \bibnamefont{and}
  \bibinfo{author}{\bibfnamefont{P.}~\bibnamefont{Schmelcher}},
  \bibinfo{journal}{Phys. Rev. Lett.} \textbf{\bibinfo{volume}{113}},
  \bibinfo{pages}{050403} (\bibinfo{year}{2014}).

\bibitem[{\citenamefont{Kalozoumis et~al.}(2015)\citenamefont{Kalozoumis,
  Morfonios, Diakonos, and
  Schmelcher}}]{Kalozoumis2015_AP_362_684_InvariantCurrents}
\bibinfo{author}{\bibfnamefont{P.~A.} \bibnamefont{Kalozoumis}},
  \bibinfo{author}{\bibfnamefont{C.~V.} \bibnamefont{Morfonios}},
  \bibinfo{author}{\bibfnamefont{F.~K.} \bibnamefont{Diakonos}},
  \bibnamefont{and}
  \bibinfo{author}{\bibfnamefont{P.}~\bibnamefont{Schmelcher}},
  \bibinfo{journal}{Annals of Physics} \textbf{\bibinfo{volume}{362}},
  \bibinfo{pages}{684} (\bibinfo{year}{2015}).

\bibitem[{\citenamefont{Zampetakis et~al.}(2016)\citenamefont{Zampetakis,
  Diakonou, Morfonios, Kalozoumis, Diakonos, and
  Schmelcher}}]{Zampetakis2016_JPAMT_49_195304_InvariantCurrent}
\bibinfo{author}{\bibfnamefont{V.~E.} \bibnamefont{Zampetakis}},
  \bibinfo{author}{\bibfnamefont{M.~K.} \bibnamefont{Diakonou}},
  \bibinfo{author}{\bibfnamefont{C.~V.} \bibnamefont{Morfonios}},
  \bibinfo{author}{\bibfnamefont{P.~A.} \bibnamefont{Kalozoumis}},
  \bibinfo{author}{\bibfnamefont{F.~K.} \bibnamefont{Diakonos}},
  \bibnamefont{and}
  \bibinfo{author}{\bibfnamefont{P.}~\bibnamefont{Schmelcher}},
  \bibinfo{journal}{J. Phys. A: Math. Theor.} \textbf{\bibinfo{volume}{49}},
  \bibinfo{pages}{195304} (\bibinfo{year}{2016}).

\bibitem[{\citenamefont{Morfonios et~al.}(2017)\citenamefont{Morfonios,
  Kalozoumis, Diakonos, and
  Schmelcher}}]{Morfonios2017_AP_385_623_NonlocalDiscrete}
\bibinfo{author}{\bibfnamefont{C.~V.} \bibnamefont{Morfonios}},
  \bibinfo{author}{\bibfnamefont{P.~A.} \bibnamefont{Kalozoumis}},
  \bibinfo{author}{\bibfnamefont{F.~K.} \bibnamefont{Diakonos}},
  \bibnamefont{and}
  \bibinfo{author}{\bibfnamefont{P.}~\bibnamefont{Schmelcher}},
  \bibinfo{journal}{Annals of Physics} \textbf{\bibinfo{volume}{385}},
  \bibinfo{pages}{623} (\bibinfo{year}{2017}).

\bibitem[{\citenamefont{R\"{o}ntgen et~al.}(2017)\citenamefont{R\"{o}ntgen,
  Morfonios, Diakonos, and
  Schmelcher}}]{Roentgen2017_AP_380_135_Non-localCurrents}
\bibinfo{author}{\bibfnamefont{M.}~\bibnamefont{R\"{o}ntgen}},
  \bibinfo{author}{\bibfnamefont{C.~V.} \bibnamefont{Morfonios}},
  \bibinfo{author}{\bibfnamefont{F.~K.} \bibnamefont{Diakonos}},
  \bibnamefont{and}
  \bibinfo{author}{\bibfnamefont{P.}~\bibnamefont{Schmelcher}},
  \bibinfo{journal}{Annals of Physics} \textbf{\bibinfo{volume}{380}},
  \bibinfo{pages}{135} (\bibinfo{year}{2017}).

\bibitem[{\citenamefont{Schmelcher et~al.}(2017)\citenamefont{Schmelcher,
  Kr\"{o}nke, and Diakonos}}]{Schmelcher2017_JCP_146_044116_DynamicsLocal}
\bibinfo{author}{\bibfnamefont{P.}~\bibnamefont{Schmelcher}},
  \bibinfo{author}{\bibfnamefont{S.}~\bibnamefont{Kr\"{o}nke}},
  \bibnamefont{and} \bibinfo{author}{\bibfnamefont{F.~K.}
  \bibnamefont{Diakonos}}, \bibinfo{journal}{J. Chem. Phys.}
  \textbf{\bibinfo{volume}{146}}, \bibinfo{pages}{044116}
  (\bibinfo{year}{2017}).

\bibitem[{\citenamefont{Wulf et~al.}(2016)\citenamefont{Wulf, Morfonios,
  Diakonos, and Schmelcher}}]{Wulf2016_PRE_93_052215_ExposingLocal}
\bibinfo{author}{\bibfnamefont{T.}~\bibnamefont{Wulf}},
  \bibinfo{author}{\bibfnamefont{C.~V.} \bibnamefont{Morfonios}},
  \bibinfo{author}{\bibfnamefont{F.~K.} \bibnamefont{Diakonos}},
  \bibnamefont{and}
  \bibinfo{author}{\bibfnamefont{P.}~\bibnamefont{Schmelcher}},
  \bibinfo{journal}{Phys. Rev. E} \textbf{\bibinfo{volume}{93}},
  \bibinfo{pages}{052215} (\bibinfo{year}{2016}).

\bibitem[{\citenamefont{Barrett et~al.}(2017)\citenamefont{Barrett, Francis,
  and Webb}}]{Barrett2017_LAA_513_409_EquitableDecompositions}
\bibinfo{author}{\bibfnamefont{W.}~\bibnamefont{Barrett}},
  \bibinfo{author}{\bibfnamefont{A.}~\bibnamefont{Francis}}, \bibnamefont{and}
  \bibinfo{author}{\bibfnamefont{B.}~\bibnamefont{Webb}},
  \bibinfo{journal}{Lin. Alg. Appl.} \textbf{\bibinfo{volume}{513}},
  \bibinfo{pages}{409} (\bibinfo{year}{2017}).

\bibitem[{\citenamefont{Francis et~al.}(2017)\citenamefont{Francis, Smith,
  Sorensen, and Webb}}]{Francis2017_LAA_532_432_ExtensionsApplications}
\bibinfo{author}{\bibfnamefont{A.}~\bibnamefont{Francis}},
  \bibinfo{author}{\bibfnamefont{D.}~\bibnamefont{Smith}},
  \bibinfo{author}{\bibfnamefont{D.}~\bibnamefont{Sorensen}}, \bibnamefont{and}
  \bibinfo{author}{\bibfnamefont{B.}~\bibnamefont{Webb}},
  \bibinfo{journal}{Lin. Alg. Appl.} \textbf{\bibinfo{volume}{532}},
  \bibinfo{pages}{432} (\bibinfo{year}{2017}).

\bibitem[{\citenamefont{Fritscher and
  Trevisan}(2016)}]{Fritscher2016_SJMAA_37_260_ExploringSymmetries}
\bibinfo{author}{\bibfnamefont{E.}~\bibnamefont{Fritscher}} \bibnamefont{and}
  \bibinfo{author}{\bibfnamefont{V.}~\bibnamefont{Trevisan}},
  \bibinfo{journal}{{SIAM} J. Matrix Anal. Appl.}
  \textbf{\bibinfo{volume}{37}}, \bibinfo{pages}{260} (\bibinfo{year}{2016}).

\bibitem[{\citenamefont{Rusek et~al.}(2000)\citenamefont{Rusek, Mostowski, and
  Or\l{}owski}}]{Rusek2000_PRA_61_022704_RandomGreen}
\bibinfo{author}{\bibfnamefont{M.}~\bibnamefont{Rusek}},
  \bibinfo{author}{\bibfnamefont{J.}~\bibnamefont{Mostowski}},
  \bibnamefont{and}
  \bibinfo{author}{\bibfnamefont{A.}~\bibnamefont{Or\l{}owski}},
  \bibinfo{journal}{Phys. Rev. A} \textbf{\bibinfo{volume}{61}},
  \bibinfo{pages}{022704} (\bibinfo{year}{2000}).

\bibitem[{\citenamefont{Pinheiro et~al.}(2004)\citenamefont{Pinheiro, Rusek,
  Orlowski, and van Tiggelen}}]{Pinheiro2004_PRE_69_026605_ProbingAnderson}
\bibinfo{author}{\bibfnamefont{F.~A.} \bibnamefont{Pinheiro}},
  \bibinfo{author}{\bibfnamefont{M.}~\bibnamefont{Rusek}},
  \bibinfo{author}{\bibfnamefont{A.}~\bibnamefont{Orlowski}}, \bibnamefont{and}
  \bibinfo{author}{\bibfnamefont{B.~A.} \bibnamefont{van Tiggelen}},
  \bibinfo{journal}{Phys. Rev. E} \textbf{\bibinfo{volume}{69}},
  \bibinfo{pages}{026605} (\bibinfo{year}{2004}).

\bibitem[{\citenamefont{Christofi et~al.}(2016)\citenamefont{Christofi,
  Pinheiro, and Dal~Negro}}]{Christofi2016_OL_41_1933_ProbingScattering}
\bibinfo{author}{\bibfnamefont{A.}~\bibnamefont{Christofi}},
  \bibinfo{author}{\bibfnamefont{F.~A.} \bibnamefont{Pinheiro}},
  \bibnamefont{and}
  \bibinfo{author}{\bibfnamefont{L.}~\bibnamefont{Dal~Negro}},
  \bibinfo{journal}{Opt. Lett.} \textbf{\bibinfo{volume}{41}},
  \bibinfo{pages}{1933} (\bibinfo{year}{2016}).

\bibitem[{\citenamefont{Chalker et~al.}(2010)\citenamefont{Chalker, Pickles,
  and Shukla}}]{Chalker2010_PRB_82_104209_AndersonLocalization}
\bibinfo{author}{\bibfnamefont{J.~T.} \bibnamefont{Chalker}},
  \bibinfo{author}{\bibfnamefont{T.~S.} \bibnamefont{Pickles}},
  \bibnamefont{and} \bibinfo{author}{\bibfnamefont{P.}~\bibnamefont{Shukla}},
  \bibinfo{journal}{Phys. Rev. B} \textbf{\bibinfo{volume}{82}},
  \bibinfo{pages}{104209} (\bibinfo{year}{2010}).

\bibitem[{\citenamefont{Mukherjee et~al.}(2015)\citenamefont{Mukherjee,
  Spracklen, Choudhury, Goldman, \"{O}hberg, Andersson, and
  Thomson}}]{Mukherjee2015_PRL_114_245504_ObservationLocalized}
\bibinfo{author}{\bibfnamefont{S.}~\bibnamefont{Mukherjee}},
  \bibinfo{author}{\bibfnamefont{A.}~\bibnamefont{Spracklen}},
  \bibinfo{author}{\bibfnamefont{D.}~\bibnamefont{Choudhury}},
  \bibinfo{author}{\bibfnamefont{N.}~\bibnamefont{Goldman}},
  \bibinfo{author}{\bibfnamefont{P.}~\bibnamefont{\"{O}hberg}},
  \bibinfo{author}{\bibfnamefont{E.}~\bibnamefont{Andersson}},
  \bibnamefont{and} \bibinfo{author}{\bibfnamefont{R.~R.}
  \bibnamefont{Thomson}}, \bibinfo{journal}{Phys. Rev. Lett.}
  \textbf{\bibinfo{volume}{114}}, \bibinfo{pages}{245504}
  (\bibinfo{year}{2015}).

\bibitem[{\citenamefont{Doretto and
  Goerbig}(2015)}]{Doretto2015_PRB_92_245124_Flat-bandFerromagnetism}
\bibinfo{author}{\bibfnamefont{R.~L.} \bibnamefont{Doretto}} \bibnamefont{and}
  \bibinfo{author}{\bibfnamefont{M.~O.} \bibnamefont{Goerbig}},
  \bibinfo{journal}{Phys. Rev. B} \textbf{\bibinfo{volume}{92}},
  \bibinfo{pages}{245124} (\bibinfo{year}{2015}).

\bibitem[{\citenamefont{Plotnik et~al.}(2011)\citenamefont{Plotnik, Peleg,
  Dreisow, Heinrich, Nolte, Szameit, and
  Segev}}]{Plotnik2011_PRL_107_183901_ExperimentalObservation}
\bibinfo{author}{\bibfnamefont{Y.}~\bibnamefont{Plotnik}},
  \bibinfo{author}{\bibfnamefont{O.}~\bibnamefont{Peleg}},
  \bibinfo{author}{\bibfnamefont{F.}~\bibnamefont{Dreisow}},
  \bibinfo{author}{\bibfnamefont{M.}~\bibnamefont{Heinrich}},
  \bibinfo{author}{\bibfnamefont{S.}~\bibnamefont{Nolte}},
  \bibinfo{author}{\bibfnamefont{A.}~\bibnamefont{Szameit}}, \bibnamefont{and}
  \bibinfo{author}{\bibfnamefont{M.}~\bibnamefont{Segev}},
  \bibinfo{journal}{Phys. Rev. Lett.} \textbf{\bibinfo{volume}{107}},
  \bibinfo{pages}{183901} (\bibinfo{year}{2011}).

\bibitem[{\citenamefont{Zhang et~al.}(2013)\citenamefont{Zhang, Braak, and
  Kollar}}]{Zhang2013_PRA_87_023613_BoundStates}
\bibinfo{author}{\bibfnamefont{J.~M.} \bibnamefont{Zhang}},
  \bibinfo{author}{\bibfnamefont{D.}~\bibnamefont{Braak}}, \bibnamefont{and}
  \bibinfo{author}{\bibfnamefont{M.}~\bibnamefont{Kollar}},
  \bibinfo{journal}{Phys. Rev. A} \textbf{\bibinfo{volume}{87}},
  \bibinfo{pages}{023613} (\bibinfo{year}{2013}).

\bibitem[{\citenamefont{Trudeau}(1994)}]{Trudeau1994____IntroductionGraph}
\bibinfo{author}{\bibfnamefont{R.~J.} \bibnamefont{Trudeau}},
  \emph{\bibinfo{title}{Introduction to Graph Theory}}
  (\bibinfo{publisher}{Dover Publications}, \bibinfo{year}{1994}),
  \bibinfo{edition}{2nd} ed.

\bibitem[{\citenamefont{Th\"{u}ne}(2016)}]{Thuene2016_AC___ExploitingEquitable}
\bibinfo{author}{\bibfnamefont{M.}~\bibnamefont{Th\"{u}ne}},
  \bibinfo{journal}{{arXiv:1605.05924}}  (\bibinfo{year}{2016}).

\bibitem[{Pyt()}]{PythTB}
\bibinfo{note}{Band structure calculations were performed using the software
  package PythTB (http://www.physics.rutgers.edu/pythtb/)}.

\bibitem[{\citenamefont{Pal et~al.}(2013)\citenamefont{Pal, {K.Maiti}, and
  Chakrabarti}}]{Pal2013_EEL_102_17004_CompleteAbsence}
\bibinfo{author}{\bibfnamefont{B.}~\bibnamefont{Pal}},
  \bibinfo{author}{\bibfnamefont{S.}~\bibnamefont{{K.Maiti}}},
  \bibnamefont{and}
  \bibinfo{author}{\bibfnamefont{A.}~\bibnamefont{Chakrabarti}},
  \bibinfo{journal}{Europhys. Lett.} \textbf{\bibinfo{volume}{102}},
  \bibinfo{pages}{17004} (\bibinfo{year}{2013}).
\end{thebibliography}
\end{document}